\documentclass[11pt,twocolumn]{article}

\usepackage[utf8]{inputenc} 
\usepackage[T1]{fontenc} 
\usepackage{amsmath} 
\usepackage{amssymb} 
\usepackage{amsthm} 
\usepackage{amsfonts}
\usepackage{amstext}
\usepackage{alltt}
\usepackage{verbatim}
\usepackage{booktabs}
\usepackage{latexsym}
\usepackage{amsfonts}
\usepackage{textcomp}
\usepackage{graphicx}
\graphicspath{{Figures/}}
\usepackage{enumitem}
\usepackage{epstopdf}
\usepackage{subfig}
\usepackage{color}
\usepackage{bm}
\usepackage{longtable}
\usepackage[labelfont=bf,labelsep=period,font=small]{caption}
\usepackage[top=2cm, bottom=2cm, left=2cm, right=2cm, scale=0.75]{geometry}
\usepackage[labelfont=bf,textfont=md]{caption}
\usepackage{pifont}
\usepackage{palatino} 
\usepackage{textcomp}
\usepackage{nicefrac}
\usepackage{hyperref}

\usepackage[export]{adjustbox}

\usepackage{setspace}

\usepackage{natbib} 
\setcitestyle{square,aysep={},yysep={;}}
\newcommand{\AuGMEnT}{\texttt{AuGMEnT}}
\newcommand{\HER}{\texttt{HER}}

\newcommand{\LSTM}{\texttt{LSTM}}

\begin{document}
	
	\twocolumn[{%
		\centering
		\LARGE Multi-timescale memory dynamics\\in a reinforcement learning network \\with attention-gated memory \\[1.5em]
		\large Marco Martinolli$^\dagger$,
		Wulfram Gerstner$^\dagger$
		and Aditya Gilra$^\dagger$\\[1em]
		\normalsize
		$^\dagger$School of Computer and Communication Sciences,\\ and Brain-Mind Institute, School of Life Sciences,\\ \'{E}cole Polytechnique F\'{e}d\'{e}rale de Lausanne, 1015 Lausanne EPFL, Switzerland \\
		Correspondence: \href{mailto:marco.martinolli@epfl.ch}{marco.martinolli@epfl.ch}, \href{mailto:aditya.gilra@epfl.ch}{aditya.gilra@epfl.ch}\\[2em]
	\begin{flushleft}
	\begin{abstract}
	 {\small 
	 Learning and memory are intertwined in our brain and their relationship is at the core of several recent neural network models. In particular, the Attention-Gated MEmory Tagging model (\AuGMEnT{}) is a reinforcement learning network with an emphasis on biological plausibility of memory dynamics and learning. We find that the \AuGMEnT{} network does not solve some hierarchical tasks, where higher-level stimuli have to be maintained over a long time, while lower-level stimuli need to be remembered and forgotten over a shorter timescale. To overcome this limitation, we introduce hybrid \AuGMEnT{}, with leaky or short-timescale and non-leaky or long-timescale units in memory, that allow to exchange lower-level information while maintaining higher-level one, thus solving both hierarchical and distractor tasks.
	}
	\end{abstract}
	\\[3em]
	\end{flushleft}
}]

\section{Introduction}
\label{introduction}
Memory spans various timescales and plays a crucial role in human and animal learning \citep{intro1}. In cognitive neuroscience, the memory system that enables manipulation and storage of information over a period of a few seconds is called Working Memory (WM), and is correlated with activity in prefrontal cortex (PFC) and basal ganglia (BG) \citep{frank,mink}. In computational neuroscience, there are not only several standalone models of WM dynamics \citep{barak,samsonovich,compte}, but also supervised and reinforcement learning models augmented by working memory \citep{HER2015,AuGMEnT,NTM,DNC,oneshotNTM}.\\ 

Memory mechanisms can be implemented by enriching a subset of artificial neurons with slow time constants and gating mechanisms \citep{LSTM,LSTM2,GRU}. More recent memory-augmented neural network models like Neural Turing Machine \citep{NTM} and Differentiable Neural Computer \citep{DNC}, employ an addressable memory matrix that works as a repository of past experiences and a neural controller that is able to store and retrieve information from the external memory to improve its learning performance.\\

Here, we study and extend the Attention-Gated MEmory Tagging model or \AuGMEnT{} \citep{AuGMEnT}. \AuGMEnT{} is trained with a Reinforcement Learning (RL) scheme, where learning is based on a reward signal that is released after each response selection. The representation of stimuli is accumulated in the memory states and the memory is reset at the end of each trial (see Methods). The main advantage of the \AuGMEnT{} network for the computational neuroscience community resides in the biological plausibility of its learning algorithm. \\

Notably, the \AuGMEnT{} network \citep{AuGMEnT} uses a memory-augmented version of a biologically plausible learning rule \citep{AGREL} mimicking backpropagation (BP). Learning is the result of the joint action of two factors, neuromodulation and attentional feedback, both influencing synaptic plasticity. The former is a global reward-related signal that is released homogeneously across the network to inform each synapse of the reward prediction error after response selection \citep{brainsigma1,brainsigma2,brainsigma3}. Neuromodulators such as dopamine influence synaptic plasticity \citep{yagishita,he,brzosko1,brzosko2,fremaux}. The novelty of \AuGMEnT{} compared to three-factor rules \citep{xie,legenstein,vasilaki,fremaux} is to add an attentional feedback system in order to keep track of the synaptic connections that cooperated for the selection of the winning action and overcome the so-called structural credit assignment problem \citep{AGREL,AuGMEnT}. \AuGMEnT{} includes a memory system, where units accumulate activity across several stimuli in order to solve temporal credit assignment tasks involving delayed reward delivery \citep{tempproblem,memoryrole}. The attentional feedback mechanism in \AuGMEnT{} works with: a) synaptic eligibility traces that decay slowly over time, and b) non-decaying neuronal traces that store the history of stimuli presented to the network up to the current time \citep{AuGMEnT} \citep{AuGMEnT}. The \AuGMEnT{} network solves the Saccade-AntiSaccade task \citep{AuGMEnT}, which is equivalent to a temporal XOR task \citep{XOR} (see Supplementary Material).\\

However, in the case of more complex tasks with long trials and multiple stimuli, like 12AX \citep{PBWM} depicted in Figure \ref{fig:AugMEnT}A, we find that the accumulation of information in \AuGMEnT{} can lead to memory interference and loss in performance. Hence, we ask the question whether a modified \AuGMEnT{} model would lead to a broader applicability of attention-gated reinforcement learning. We propose a variant of the \AuGMEnT{} network, named hybrid \AuGMEnT{}, that introduces timescales of forgetting or leakage in the memory dynamics to overcome this kind of learning limitation. We employ memory units with different decay constants so that they work on different temporal scales, while the network learns to weight their usage based on the requirements of the specific task. In our simulations, we employed just two subgroups of cells in the memory, where one half of the memory is non-leaky and the other is leaky with a uniform decay time constant; however, more generally, the hybrid \AuGMEnT{} architecture may contain several subgroups with distinct leakage behaviours.\\

The paper is structured as follows. Section \ref{methods} presents the architectural and mathematical details of hybrid \AuGMEnT{}. Section \ref{results} describes the simulated results of the hybrid \AuGMEnT{} network, the standard \AuGMEnT{} network and a fully leaky control network, on two cognitive tasks, a non-hierarchical task involving sequence prediction \citep{seq_pred} and a hierarchical task 12AX \citep{PBWM}. Finally, in Section \ref{discussion} we discuss our main achievements in comparison with state-of-the-art models and present possible future developments of the work.\\

\section{Methods}
\label{methods}

\subsection{Hybrid \AuGMEnT{} network architecture and operation}

\begin{figure*}[h!] 
	\begin{minipage}[t]{0.03\textwidth}
        \textbf{A}
    \end{minipage}
    \begin{minipage}[t]{0.45\textwidth}
	    \includegraphics[width = 0.9\textwidth,valign=t]{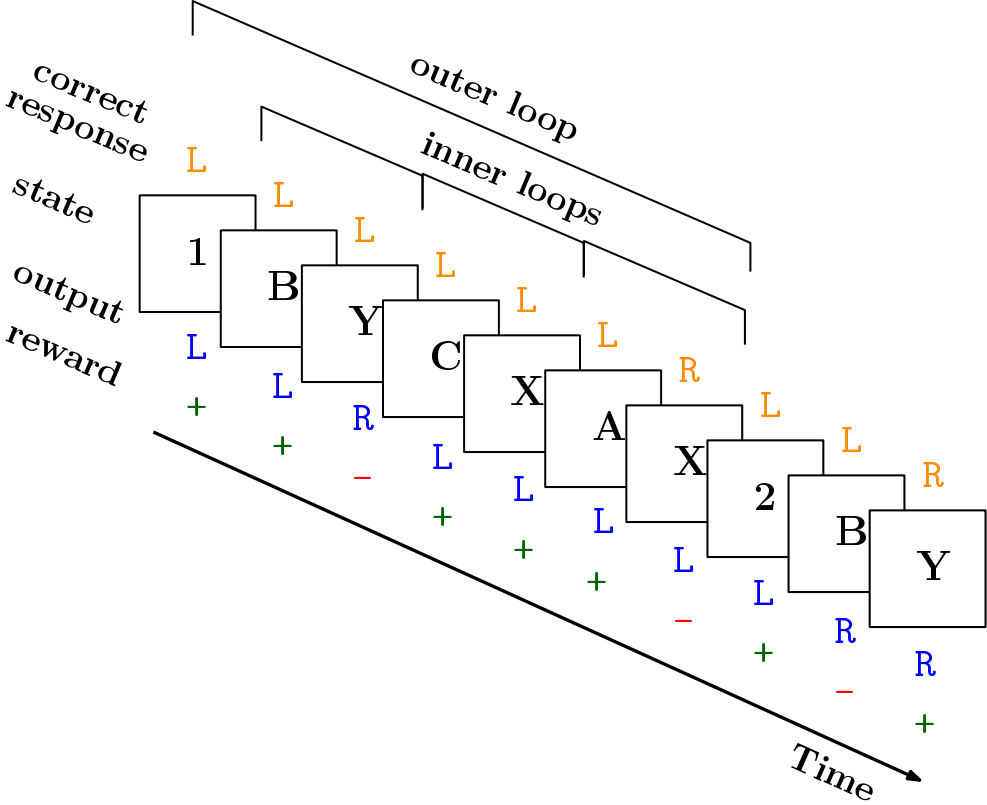}
    \end{minipage}\hfill
	\begin{minipage}[t]{0.03\textwidth}
        \textbf{B}
    \end{minipage}
    \begin{minipage}[t]{0.45\textwidth}
    	\bigskip
	    \includegraphics[width = 0.9\textwidth,valign=t]{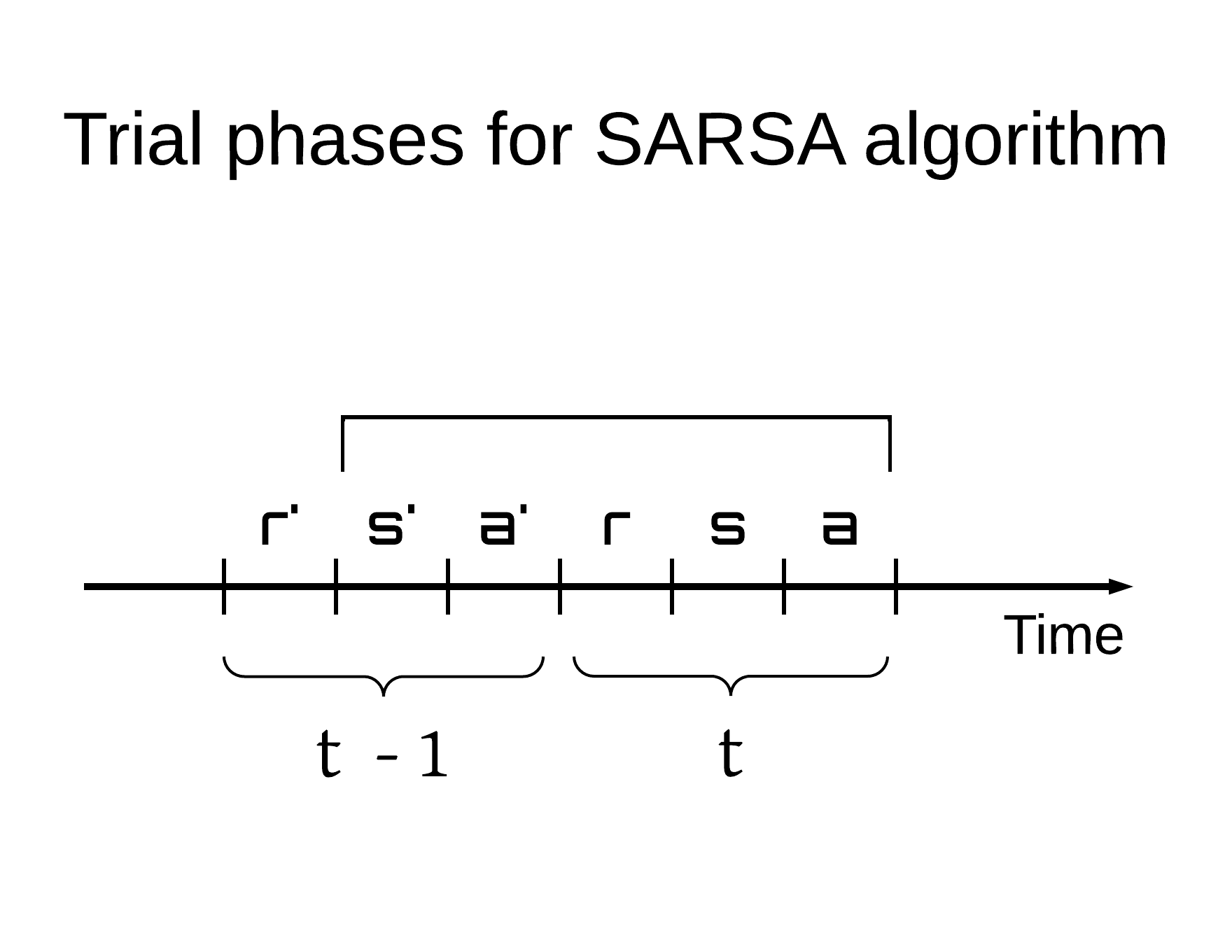}
    \end{minipage}\hfill
    \\[0.75cm]
	\begin{minipage}[t]{0.03\textwidth}
        \textbf{C}
    \end{minipage}
    \begin{minipage}[t]{0.93\textwidth}
	    \includegraphics[width = 0.495\textwidth,valign=t]{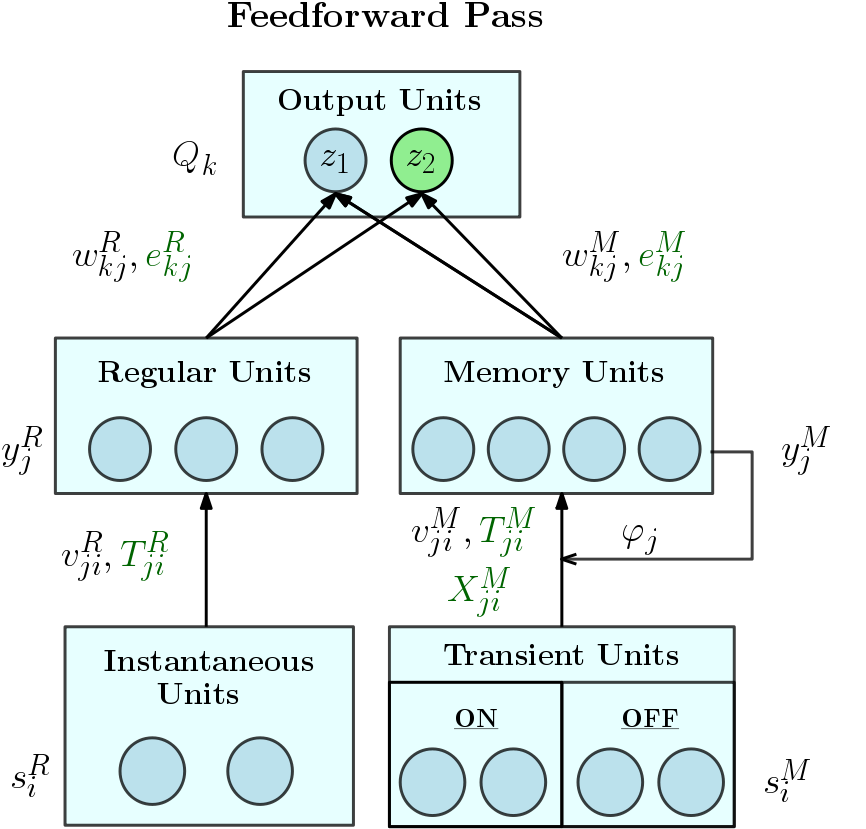}
        \qquad
		\includegraphics[width = 0.495\textwidth,valign=t]{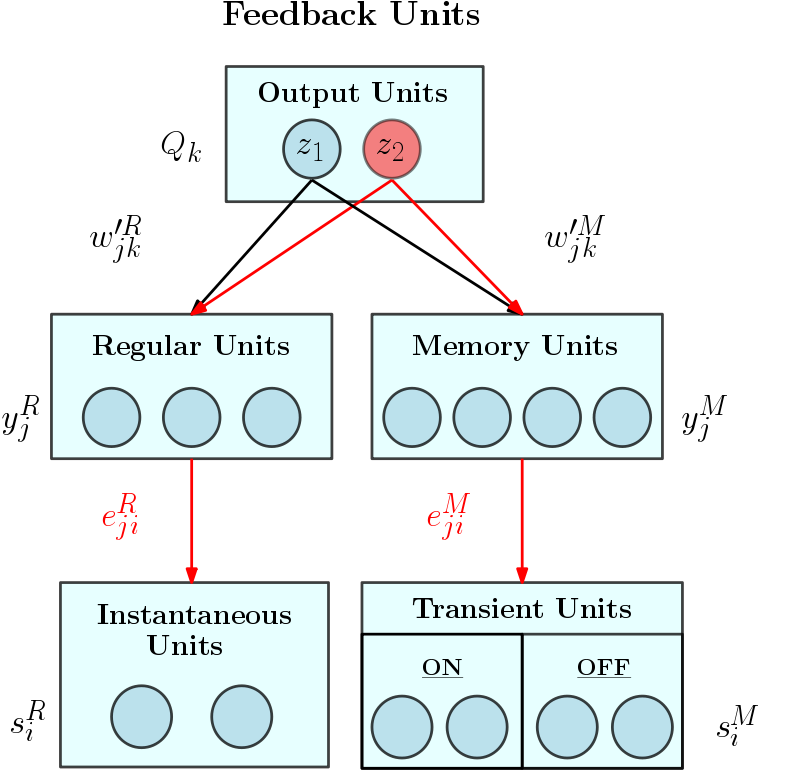}
    \end{minipage}\hfill     
	\caption[]{\textbf{Overview of \AuGMEnT{} network operation.} \textbf{A}. Example of trials in the 12AX task, where task symbols appear sequentially on a screen organized in outer loops, which start with either digit \texttt{1} or \texttt{2}, and a random number of inner loops (e.g. \texttt{B}-\texttt{Y}, \texttt{C}-\texttt{X} and \texttt{A}-\texttt{X}). Each cue presentation is associated with a Target (\texttt{R}) or Non-Target (\texttt{L}) correct response. When output and correct response coincide, the agent receives a positive reward (+), otherwise it gets a negative reward (-). Figure is adapted from Figure 1 of \cite{PBWM}. \textbf{B}. \AuGMEnT{} operates in discrete time steps each comprising the reception of reward (r), input of state or stimulus (s) and action taken (a). It implements the State-Action-Reward-State-Action (SARSA, in figure s'a'rsa) reinforcement learning algorithm. In time step $t$, reward r is obtained for the previous action a' taken in time step $t-1$. The network weights are updated once the next action a is chosen. \textbf{C}. The \texttt{AuGMEnT} network is structured in three layers with different types of units. Each iteration of the learning process consists of a feedforward pass (left) and a feedback pass (right). In the feedforward pass, sensory information about the current stimulus in the bottom layer, is fed to regular units without memory (left branch) and units with memory (right branch) in the middle layer, whose activities in turn are weighted to compute the Q-values in the top activity layer. Based on the Q-values, the current action is selected (e.g. green $z_2$). The reward obtained for the previous action is used to compute the reward prediction error, which modifies the connection weights, that contributed to the selection of the previous action, in proportion to their eligibility traces. After this, temporal eligibility traces and tags (in green) on the connections are updated to reflect the correlations between the current pre and post activities. Then, in the feedback pass, spatial eligibility traces (in red) are updated, attention-gated by the current action (e.g. red $z_2$), via feedback weights. }
	\label{fig:AugMEnT}
\end{figure*}

The network controls an agent which, in each time step $t$, receives a reward in response to the previous action, processes the next stimulus, and takes the next action, as in Figure \ref{fig:AugMEnT}B. In each time step, we distinguish two phases, called the feedforward pass and feedback pass, depicted in Figure \ref{fig:AugMEnT}C.

\subsubsection{Feedforward pass: stimulus to action selection}

In \texttt{AuGMEnT} \citep{AuGMEnT}, information is processed through a network with three layers, as shown in the left panel of Figure \ref{fig:AugMEnT}C. Each unit of the output layer corresponds to an action. There are two pathways into the output layer: the regular $R$ branch and the memory $M$ branch.\\

The regular branch is a standard feedforward network with one hidden layer. The current stimulus $s^R_i(t)$, indexed by unit index $i=1,\ldots,S$ is connected to the hidden units (called regular units) indexed by $j$, via a set of modifiable synaptic weights $v^{R}_{ji}$ yielding activity $y^{R}_j$:
\begin{equation}
y_j^{R}(t) = \sigma\left( h^R_j \right),\quad h^R_j = \sum_i v_{ji}^{R} s^R_i(t),
\label{AuGMEnT_regular_layer}
\end{equation}
where $\sigma$ is the sigmoidal function $\sigma(x)=(1+\exp(-x))^{-1}$. Input units are one-hot binary with values $S_i \in \{0,1\}$ (equal to 1 if stimulus $i$ is currently presented, 0 otherwise).\\	

\begin{figure*}[h!]
		\centering
		\begin{minipage}[t]{0.45\textwidth}
				\quad \quad \quad \quad \quad \quad \, \,
					{\Large \AuGMEnT{}}
				\vskip 5mm
		    	\includegraphics[width = \textwidth]{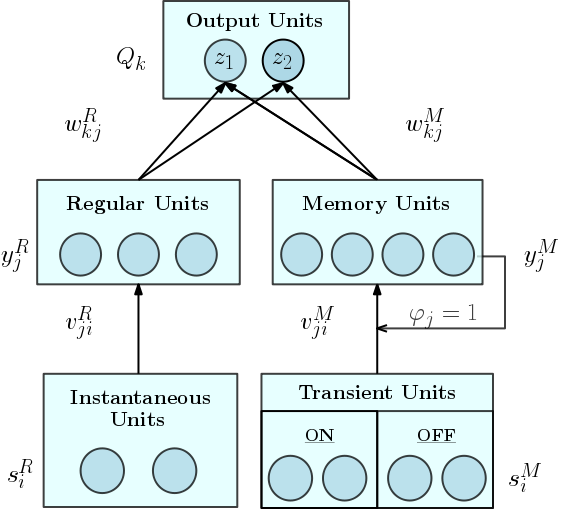}
		\end{minipage}\hfill
		\begin{minipage}[t]{0.45\textwidth}
				\quad \quad \quad \quad \,
                {\Large Hybrid \AuGMEnT{}}
				\vskip 3.8mm			
			\includegraphics[width = \textwidth]{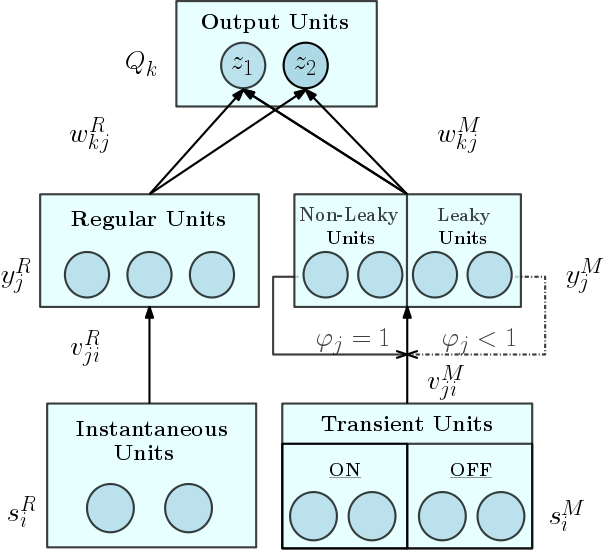}
		\end{minipage}\hfill
		\caption[]{\textbf{Architectures of standard \AuGMEnT{} and hybrid \AuGMEnT{} networks.} The difference between the networks consist in their memory dynamics: the memory layer of standard \AuGMEnT{} (left) has only conservative units with $\varphi_j\equiv 1$, while hybrid \AuGMEnT{} (right) possesses a memory composed of both leaky $\varphi_j<1$ and non-leaky $\varphi_j=1$ units.}
		\label{fig:AugMEnT_var}
	\end{figure*}

The memory branch is driven by \emph{transitions} between stimuli, instead of the stimuli themselves. The sensory input of the memory branch consists of a set of $2S$ transient units,  i.e. $S$ ON units $s_l^{+} \in \left\lbrace 0,1\right\rbrace^{S}$ that encode the onset of each stimulus, and $S$ OFF units $s_l^{-} \in \left\lbrace 0,1\right\rbrace ^{S}$ that encode the offset:
\begin{align}
\begin{split}
s_l^{+}(t) &= [ s_l(t)-s_l(t-1) ]_{+} \\
s_l^{-}(t) &= [ s_l(t-1)-s_l(t) ]_{+},
\end{split}
\end{align}
where the brackets signify rectification. In the following, we denote the input into the memory branch with a variable $s_i^M$ defined as the concatenation of these ON and OFF units:
\begin{align}
s^M_i(t) &= \begin{cases} 
 s_i^{+}(t), &\textnormal{ if } i\le S \\
 s_{i-S}^{-}(t), &\textnormal{ if  } i>S, \end{cases}
\label{AuGMEnT_transient_stimuli}
\end{align}
The memory units have to maintain task-relevant information through time. The transient input is transmitted via the synaptic connections $v_{ji}^{M}$ to the memory layer, where it is accumulated in the states:
\begin{equation}
h^{M}_j(t) = \varphi_j h^{M}_j(t-1) + \sum_i v_{ji}^{M} s_i^{M}(t).
\label{AuGMEnT_cumulative_memory}
\end{equation}
We introduce the factor $\varphi_j \in [0,1]$ here, as an extension to the standard \AuGMEnT{} \citep{AuGMEnT}, to incorporate decay or forgetting of the memory state $h_j^M$ over time. Setting $\varphi_j \equiv 1$ for all $j$, we obtain non-leaky memory dynamics as in the original \AuGMEnT{} network \citep{AuGMEnT} (Fig. \ref{fig:AugMEnT_var}, left panel). In our hybrid \AuGMEnT{} network, each memory cell or subgroup of memory cells may be assigned different leak co-efficients $\varphi_j$ (Fig. \ref{fig:AugMEnT_var}, right panel). In this way, the memory is composed of subpopulations of neurons that cooperate in different ways to solve a task, allowing at the same time long-time maintenance and fast decay of information in memory. In contrast to the forget gate of Long Short-Term Memory \citep{LSTM} or Gated Recurrent Unit \citep{GRU}, our memory leak co-efficient is not trained and gated, but fixed.\\

The memory state $h^M_j$ leads to the activation of a memory unit:
\begin{equation}
y_j^{M}(t) = \sigma\left( h_j^{M}(t)\right).
\label{AuGMEnT_memory_layer}
\end{equation}
The states of the memory units are reset to $0$ at the end of each trial.\\

Both branches converge onto the output layer. The activity of an output unit with index $k$ approximates the Q-value of action $k=a$ given the input $\mathbf{s}\equiv\lbrack s_i\rbrack$, denoted as $Q^{\mathbf{s},a}(t)$. Q-values are formally defined as the future expected discounted reward conditioned on stimulus $\textbf{s}(t)$ and action $a(t)$, that is:
\begin{equation}
Q^{\textbf{s},a}(t) = E\left[ \sum_{\tau=0}^{\infty}{\gamma^{\tau} r_{t+\tau+1}}\,\bigg|\,\textbf{s}=\textbf{s}(t),\,a=a(t)\right] 
\label{AuGMEnT_Q-value}
\end{equation}
where $\gamma \in [0,1]$ is a discount factor. Numerically, the vector $\textbf{Q}$ that approximates the Q-values is obtained by combining linearly the hidden states from the regular and the memory branches, with synaptic weights $w_{kj}^{R}$ and $w_{kj}^{M}$:
\begin{equation}
Q_k(t) = \sum_j w_{kj}^{R} y^R_j(t) + \sum_j w_{kj}^{M} y^M_j(t).
\label{AuGMEnT_qvalue}
\end{equation}
Finally, the Q-values of the different actions participate in an $\epsilon$-greedy winner-take-all competition to select the response of the network. With probability $1-\epsilon$, the next action $a(t)$ is the one with the maximal $Q$-value:
\begin{equation}
a(t) = \textnormal{argmax}_k Q_k(t).
\end{equation}
With probability $\epsilon$, a stochastic policy is chosen with probability:
\begin{equation}
p_{a} = \dfrac{\exp(g(t)Q_{a})}{\sum_{k}{\exp(g(t)Q_{k})}}
\label{AuGMEnT_softmax}
\end{equation}
where $g(t)$ is a weight function defined as $g(t) = 1 + \frac{10}{\pi}\arctan(\frac{t}{t^{*}})$, that gradually increases in time over a task-specific, fixed time scale $t^{*}$. Over time, this emphasizes the action with maximal $Q$-value, improving prediction stability.\\

\subsubsection{Feedforward pass: reward-based update of weights, and correlation-based update of eligibility traces and tags}
\AuGMEnT{} follows the SARSA updating scheme and updates the $Q$-values for the previous action $a'$ taken at time $t-1$, once the action $a$ at time $t$ is known (see Fig. \ref{fig:AugMEnT}B). $Q$-values depend on the weights via equation \eqref{AuGMEnT_qvalue}. The temporal difference (TD) error is defined as \citep{Wiering1998,sutton1998}:
\begin{equation}
\delta(t) = \left( r(t) + \gamma Q_{a}(t)\right)- Q_{a'}(t-1),
\label{AuGMEnT_delta}
\end{equation}
where $a$ is the action chosen at current time $t$, and $r(t)$ is the reward obtained for the action $a'$ taken at time $t-1$. The Temporal Difference (TD) error $\delta(t)$ acts as a global reinforcement signal to modify the weights of all connections as
\begin{equation}
\begin{split}
v_{ji}^{R,M}(t+1) &= v_{ji}^{R,M}(t) + \beta e_{ji}^{R,M}(t) \delta(t), \\
w_{kj}^{R,M}(t+1) &= w_{kj}^{R,M}(t) + \beta e_{kj}^{R,M}(t) \delta(t),
\label{AuGMEnT_weights_update}
\end{split}
\end{equation}
where $\beta$ is a learning rate and $e_{ji}^{R,M}$ and $e_{kj}^{R,M}$ are synaptic eligibility traces, defined below. Superscript R or M denotes the regular or memory branch respectively. We use the same symbol $e^{R,M}$ for eligibility traces at the input-to-hidden ($i$ to $j$) and hidden-to-output ($j$ to $k$) synapses, even though these are different, with the appropriate one clear from context and the convention for indices.\\

After the update of weights, a synapse from neuron $j$ in the hidden layer to neuron $k$ in the output layer updates its temporal eligibility trace
\begin{equation}
\begin{split}
e^{R}_{kj}(t+1) &= y^{R}_j(t) z_k(t) + (1-\alpha) e^{R}_{kj}(t), \\
e^{M}_{kj}(t+1) &= y^{M}_j(t) z_k(t) + (1-\alpha) e^{M}_{kj}(t),
\label{AuGMEnT_tagging_1}
\end{split}
\end{equation}
where $\alpha \in [0,1]$ is a decay parameter, $z_{k}$ is a binary one-hot variable that indicates the winning action (equal to $1$ if action $k$ has been selected, $0$ otherwise), and $M$ or $R$ denotes the regular or the memory branch respectively.\\

Similarly, a synapse from neuron $i$ in the input layer to neuron $j$ in the hidden layer sets momentary tags $T_{ji}^{R,M}$ as:
\begin{equation}
\begin{split}
T_{ji}^{R}(t) &= s^{R}_i(t) \, \sigma'(h^{R}_j(t)), \\
T_{ji}^{M}(t) &= X^{M}_{ji}(t) \, \sigma'(h^{M}_j(t)),
\label{AuGMEnT_tagging_2}
\end{split}
\end{equation}
where $\sigma'(h^{R,M}_j)$ is a nonlinear function of the input potential, defined as the derivative of the gain function $\sigma$, and $X_{ji}^M$ is a synaptic trace \citep{pfister,morrison} defined as follows:
\begin{equation}
X^{M}_{ji}(t) = \varphi_{j}\,X^{M}_{ji}(t-1) + s_i^{M}(t).
\label{AuGMEnT_syntrace}
\end{equation}
Note that the tag $T_{ji}^{R,M}$ has no memory beyond one time step, i.e. it is set anew at each time step. Nevertheless, since $X_{ji}^M$ depends on previous times, the tag $T_{ji}^M$ of memory units can link across time steps. Since activities $y^{R,M}_j$, $z_k$, $s_i^{R,M}$  and input potentials $h_j^{R,M}$ are quantities available at the synapse, a biological synapse can implement the updates of eligibility traces and tags locally. We emphasize that both eligibility traces and tags can be interpreted as 'Hebbian' correlation detectors. In the original \AuGMEnT{} model \citep{AuGMEnT}, all eligibility traces and tags were said to be updated in the feedback pass. Here, without changing the order of operations of the algorithm, we have conceptually shifted the update of those traces and tags that depend on the correlations of the activities, to the last step of the feedforward pass when these activities are still available. Note that activities could in principle change via attention-gating during the feedback pass \citep{gateattention1,gateattention2}.

\subsubsection{Feedback pass: attention-gated update of eligibility traces}
After action selection and the updates of weights, tags, and temporal eligibility traces in the feedforward pass, the synapses that contributed to the currently selected action update their spatial eligibility traces in an attentional feedback step. For the synapses from the input to the hidden layer, the tag $T^{R,M}_{ji}$ from equation \eqref{AuGMEnT_tagging_2} is combined with a spatial eligibility trace which can be interpreted as an attentional feedback signal \citep{AuGMEnT}.
\begin{equation}
\begin{split}
e^{R}_{ji}(t+1) &= T_{ji}^{R} \sum_k {w'}_{jk}^{R} z_k + (1-\alpha) e^{R}_{ji}(t), \\
e^{M}_{ji}(t+1) &= T_{ji}^{M} \sum_k {w'}_{jk}^{M} z_k + (1-\alpha) e^{M}_{ji}(t),
\label{AuGMEnT_tagging_3}
\end{split}
\end{equation}
where feedback weights from the output layer to the hidden layer have been denoted as $w'_{jk}$ and $z_k \in \{0,1\}$ is the value of output unit $k$ (one-hot response vector as defined for equation \eqref{AuGMEnT_tagging_1}).\\

It must be noted that the feedback synapses ${w'}_{jk}^{R,M}$ follow the same update rule as their feedforward partner $w_{kj}^{R,M}$. Therefore, even if the initializations of the feedforward and feedback weights are different, their strengths become similar during learning, as suggested by neurophysiological findings \citep{mao2011}.\\

\subsection{Deriving the learning rules via gradient descent}

For networks with one hidden layer and one-hot coding in the output, attentional feedback is equivalent to backpropagation \citep{AGREL,AuGMEnT}. We can show that the equations for eligibility traces and tagging along with the weight update equations reduce an RPE-based loss function $E$ defined as:
\begin{equation}
E=\frac{1}{2} \left(\delta(t)\right)^2
\label{AuGMEnT_functional}
\end{equation}
Here we specifically discuss the case of the tagging equations \eqref{AuGMEnT_tagging_2} and \eqref{AuGMEnT_tagging_3} and the update rule \eqref{AuGMEnT_weights_update} associated with the weight $v_{ji}^{M}$ from sensory input into memory, as it contains the memory decay factor $\varphi_j$ that we introduced, but analogous discussion holds also for weights $v_{ji}^{R}$, $w_{kj}^{M}$ and $w_{kj}^{R}$.\\  
\begin{proof}
	We want to show that 
	\begin{equation}
		\Delta v_{ji}^{M} = \beta \,e_{ji}^{M} \,\delta_{t}\propto -\dfrac{\partial E}{\partial v_{ji}^{M}}
		\label{AuGMEnT_theorem}
	\end{equation}
	For simplicity, here we prove \eqref{AuGMEnT_theorem} neglecting the temporal decay of the eligibility trace $e^M_{ji}$ (i.e. $\alpha=1$), so that
    \begin{equation*}
    e^{M}_{ji} = T_{ji}^{M} \sum_k w_{jk}^{'M} z_k = T_{ji}^{M}\,w_{ja'}^{'M}
    \end{equation*}
    where $a'$ is the selected action at time $t-1$.\\ 
    
	We first observe that the right-hand side of equation \eqref{AuGMEnT_theorem} can be rewritten as:
	\begin{equation*}
		-\dfrac{\partial E}{\partial v_{ji}^{M}} = -\dfrac{\partial E}{\partial Q_{a'}}\,\dfrac{\partial Q_{a'}}{\partial v_{ji}^{M}} = \delta_{t}\,\dfrac{\partial Q_{a'}}{\partial v_{ji}^{M}}
	\end{equation*}
	Thus, it remains to show that $\dfrac{\partial Q_{a'}}{\partial v_{ji}^{M}}=e^{M}_{ji}$.\\
    
    Similarly to the approach used in backpropagation, we now apply the chain rule and we focus on each term separately:
	\begin{equation*}
		\dfrac{\partial Q_{a'}}{\partial v_{ji}^{M}} = \dfrac{\partial Q_{a'}}{\partial y_{j}^{M}}\, \dfrac{\partial y_{j}^{M}}{\partial h_{j}^{M}}\, \dfrac{\partial h_{j}^{M}}{\partial v_{ji}^{M}}
	\end{equation*}
    From equations \eqref{AuGMEnT_memory_layer} and \eqref{AuGMEnT_qvalue}, we immediately have that:
    \begin{equation*}
    \dfrac{\partial y_{j}^{M}}{\partial h_{j}^{M}} = \sigma'(h_{j}^{M}) \qquad \dfrac{\partial Q_{a'}}{\partial y_{j}^{M}}=w_{a'j}^M
    \end{equation*}
	However, in the feedback step the weight $w_{a'j}^{M}$ is replaced by its feedback counterpart $w_{ja'}'^M$. As discussed above, this is a valid approximation because they become similar during learning.\\
	Finally, starting from equation \eqref{AuGMEnT_cumulative_memory} we can write:
	\begin{align*}
	h_{j}^{M}(t) &= \sum_{i} v_{ji}^{M}(t)\,s_{i}^{M}(t) + \sum_{\tau=t_{0}}^{t-1}\sum_{i} \varphi_j^{t-\tau} v_{ji}^{M}(\tau)\,s_{i}^{M}(\tau) \\ &\approx \sum_{i} v_{ji}^{M}(t)\,\sum_{\tau=t_{0}}^{t} \varphi_j^{t-\tau} s_{i}^{M}(\tau)
	\end{align*}
	where $t_{0}$ indicates the starting time of the trial and last approximation derives from the assumption of slow learning dynamics, i.e. $v_{ij}^{M}(\tau)=v_{ij}^{M}(t)$ for $t_{0}\leq\tau<t$. As a consequence, we have:
	\begin{equation*}
	\dfrac{\partial h_{j}^{M}(t-1)}{\partial v_{ji}^{M}(t-1)} \approx \sum_{\tau=t_{0}}^{t-1} \varphi^{t-\tau+1}_j s_{i}^{M}(\tau)=X^{M}_{ji}(t-1)
	\label{third_derivative}
	\end{equation*}
    
	In conclusion, we combine the different terms and we obtain the desired result:
	\begin{equation*}
		\Delta v_{ji}^{M} \propto \delta_{t}\, X^{M}_{ji}\,\sigma'(h_{j}^{M})\,w_{ja'}^{'M}=\delta_{t}\,e_{ji}^{M}.
	\end{equation*}	
	\end{proof}

\subsection{Simulation and tasks}
All simulation scripts were written in python (https://www.python.org/), with plots rendered using the matplotlib module (http://matplotlib.org/). These simulation and plotting scripts are available online at https://github.com/martin592/hybrid\_AuGMEnT.\\

\begin{table}[h!]
	\centering
	\begin{tabular}{l| l l}
		\toprule
		\textbf{Parameter} & \textbf{Value} \\
		\midrule
		$\beta$ : Learning parameter & $0.15$\\
		$\lambda$ : Eligibility persistence & $0.15$\\
		$\gamma$ : Discount factor & $0.9$\\
		$\alpha$ : Eligibility decay rate & $1-\gamma\lambda$\\
		$\epsilon$ : Exploration rate & $0.025$\\						
		\bottomrule
	\end{tabular}
	\caption{Parameters for the \AuGMEnT{} network.}
    \label{tab:sim_params}
\end{table}	

We use the parameters listed in Table \ref{tab:sim_params} for our simulations. Further, for the Hybrid \AuGMEnT{} network, we set $\varphi_j=1$ for the first half of the memory cells and $\varphi_j=0.7$ for the second half. To reduce to the standard \AuGMEnT{} \citep{AuGMEnT} network, we set $\varphi_j \equiv 1$ for all $j$, while for a leaky control network we set $\varphi_j \equiv 0.7$ for all $j$. In general, the leak co-efficients may be tuned to adapt the overall memory dynamics to the specific task.\\

\section{Results}
\label{results}
\AuGMEnT{} \citep{AuGMEnT} includes a differentiable memory system and is trained in an RL framework with learning rules based on the joint effect of synaptic tagging, attentional feedback and neuromodulation (see Methods). Here, we study our proposed variant of \AuGMEnT{}, named hybrid \AuGMEnT{}, that has an additional leak factor in a subset of memory units, and compare it to the original \AuGMEnT{} and to a control network with all leaky memory units.\\

\begin{table*}[h!]
	\caption{Network Architecture Parameters for the Simulations}
	\label{tab:arch_param}
	\centering
	\begin{tabular}{ |c|c|c| }
		\toprule
		\textbf{Network Parameter} & \textbf{Sequence Prediction Task} & \textbf{12AX Task} \\
		&        ($L$ = sequence length)     &  					\\
		\midrule
		$S$ : Number of sensory units & $L-1$ & $8$\\		
		$R$ : Number of regular units & $3$ & $10$\\
		$M$ : Number of memory units & $8$ & $20$\\	
		$A$ : Number of activity units & $2$ & $2$\\
		\bottomrule
	\end{tabular}
\end{table*}
As a first step, we validated our implementations of standard and hybrid \AuGMEnT{} networks on the Saccade-AntiSaccade (S-AS) task, used in the reference paper \citep{AuGMEnT} (Supplementary Material). We next simulated the networks on two other cognitive tasks with different structure and memory demands: the sequence prediction task \citep{seq_pred} and the 12AX task \citep{PBWM}. In the former, the agent has to predict the final letter of a sequence depending only on its starting letter, while in the latter, the agent has to identify target pairs inside a sequence of hierarchical symbols. The S-AS task maps to a temporal XOR task \citep{XOR}, thus the hidden layer is essential for the task \citep{xor1,xor2}. The 12AX also resembles an XOR structure, but is more complex due to an additional dimension and distractors in the inner loop (Supplementary Figure \ref{fig:spatial_tasks}). The complexity of the sequence prediction task is less compared to the 12AX task, and can be effectively solved by \AuGMEnT{}. We will show that hybrid \AuGMEnT{} performs well on both cognitive tasks, whereas standard \AuGMEnT{} fails on the 12AX task. The parameters involving the architecture of the networks on each task are reported in Table \ref{tab:arch_param}. We now discuss each of the tasks in more detail.\\

\subsection{Task 1: Sequence Prediction}	
\label{seq_pred}
\begin{figure}[h!]
	\centering
	\includegraphics[width = 0.4\textwidth]{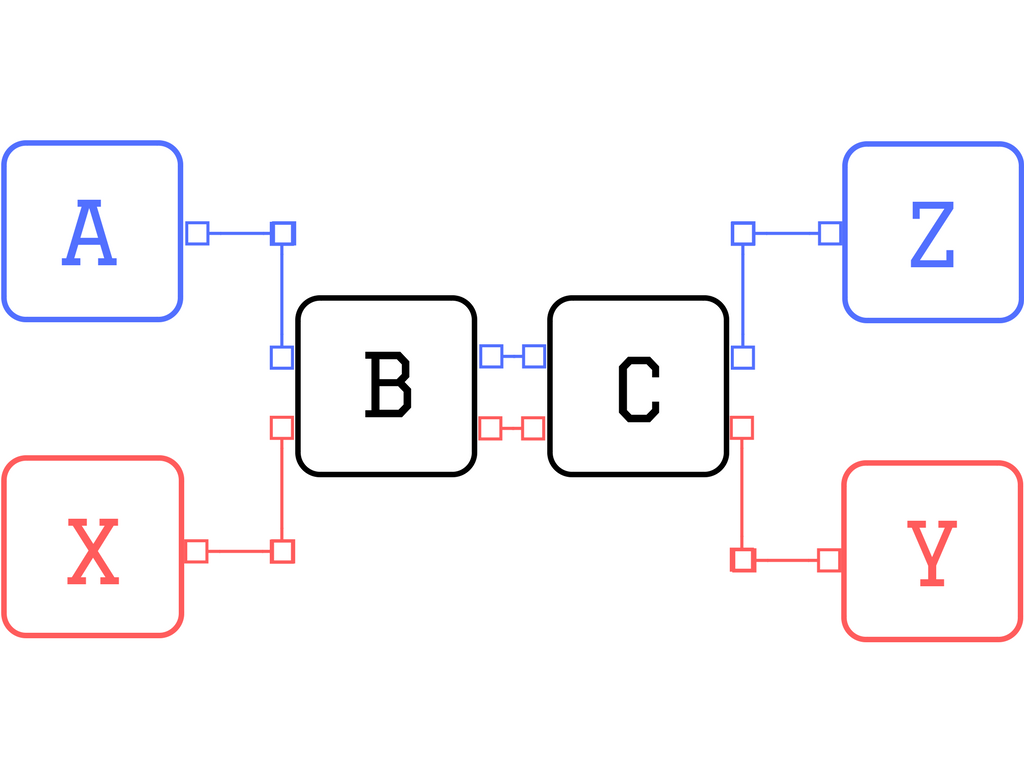}
	\caption[]{\textbf{Scheme of the sequence prediction task.} Scheme of sequence prediction trials with sequence length equal to 4 (i.e. 2 distractors): the two possible sequences are: \texttt{A}-\texttt{B}-\texttt{C}-\texttt{Z} (blue) or \texttt{X}-\texttt{B}-\texttt{C}-\texttt{Y} (red)}
	\label{fig:seq_pred_task}
\end{figure}

In the sequence prediction task \citep{seq_pred}, letters appear sequentially on a screen and at the end of each trial the agent has to correctly predict the last letter. Each sequence starts either with an \texttt{A} or with an \texttt{X}, which is followed by a fixed sequence of letters (e.g. \texttt{B}-\texttt{C}-\texttt{D}-\texttt{E}). The trial ends with the prediction of the final letter, which depends on the initial cue: if the sequence started with \texttt{A}, then the final letter has to be a \texttt{Z}; if the initial cue was an \texttt{X}, then the final letter has to be a \texttt{Y}. In case of correct prediction the agent receives a reward of $1$ unit, otherwise he is punished with a negative reward of $-1$. A scheme of the task is presented in Figure \ref{fig:seq_pred_task} for sequences of four letters.\\
The network has to learn the task for a given sequence length, kept fixed throughout training. The agent must learn to maintain in memory, the initial cue of the sequence until the end of the trial, to solve the task. At the same time, the agent has to learn to neglect the information coming from the intermediate cues (called distractors). Thus the difficulty of the task is correlated with the length of the sequence.\\

\begin{figure*}[h!]
	\centering
	\begin{minipage}[t]{0.03\textwidth}
        \textbf{A}
    \end{minipage}
    \begin{minipage}[t]{0.44\textwidth}
	    \includegraphics[width = \textwidth,valign=t]{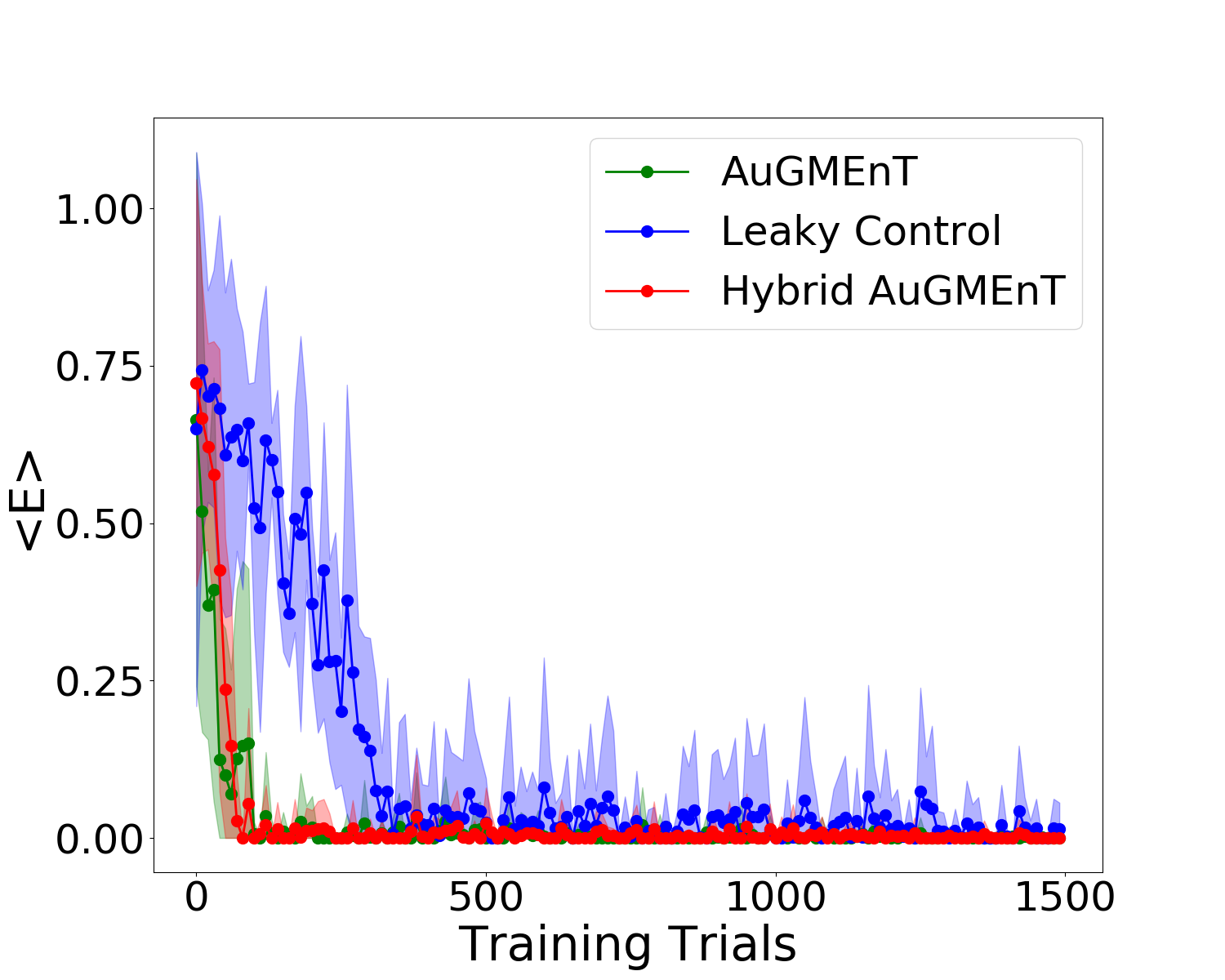}
    \end{minipage}\hfill
    \begin{minipage}[t]{0.03\textwidth}
        \textbf{B}
    \end{minipage}
    \begin{minipage}[t]{0.44\textwidth}	
	    \includegraphics[width = \textwidth,valign=t]{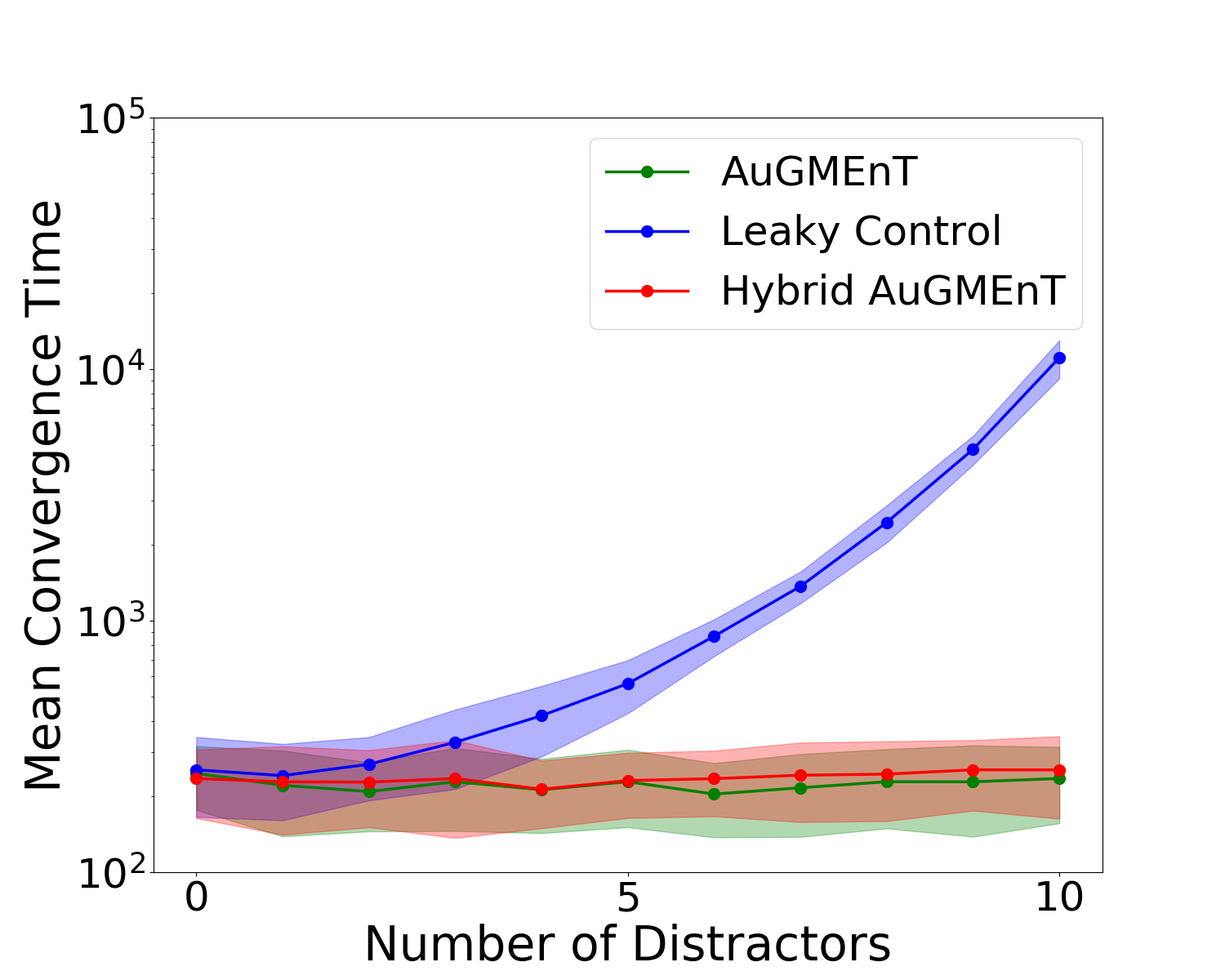}
	\end{minipage}
	\caption[]{\textbf{Convergence in the sequence prediction task.} \textbf{A}. Time course of error of the models on the sequence prediction task with sequences of five letters (three distractors): the mean squared RPE decays to zero for all networks but the leaky control network (blue) is much slower than \AuGMEnT{} (green) and Hybrid \AuGMEnT{} (red). \textbf{B}. Convergence time of the \AuGMEnT{} network and its variants on the sequence prediction task with increasing number of distractors, i.e. intermediate cues before final prediction.}
	\label{fig:distr_analysis}
\end{figure*}
\begin{figure*}[h!]
	\centering
	\includegraphics[width = \textwidth]{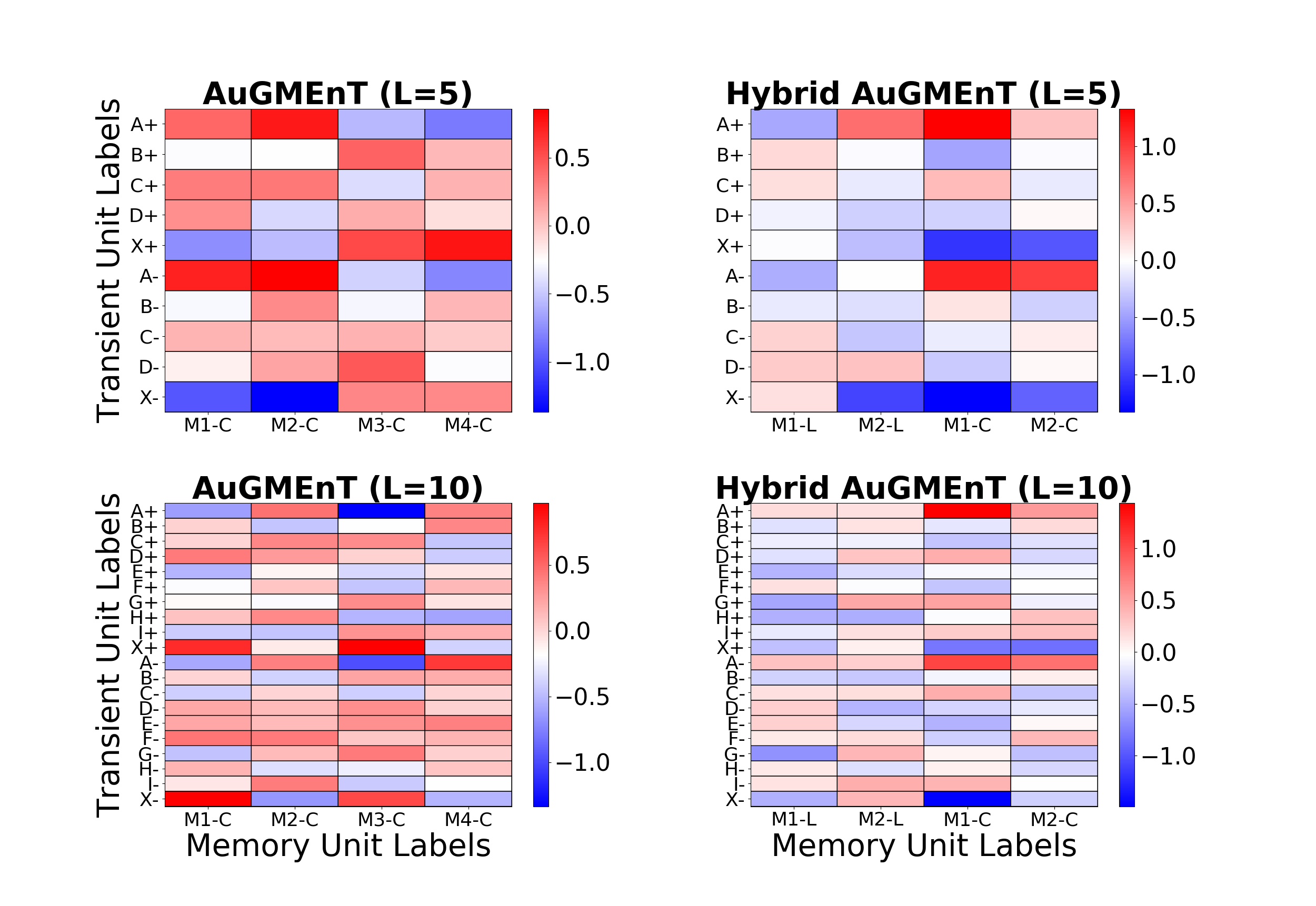}	
	\caption[]{\textbf{Memory weights of \AuGMEnT{} networks in sequence prediction task.} Memory weight matrices after convergence for \AuGMEnT{} (left) and Hybrid \AuGMEnT{} (right) networks on the sequence prediction task with sequences of length five (first row) and ten (second row). Note that the first two memory units in Hybrid \AuGMEnT{} are leaky (\texttt{M1-L} and \texttt{M2-L}), while the last ones are conservative or non-leaky (\texttt{M1-C} and \texttt{M2-C}).}
	\label{fig:weight_analysis}
\end{figure*}
\begin{table*}[h!]
	\caption{The 12AX task: table of key information}
	\label{tab:12AX_param}
	\centering
	\begin{tabular}{l l}
		\toprule
		\textbf{Task feature} & \textbf{Details}  \\
		\midrule
		Input & $8$ possible stimuli: \texttt{1},\texttt{2},\texttt{A},\texttt{B},\texttt{C},\texttt{X},\texttt{Y},\texttt{Z}.\\
		Output & Non-Target (\texttt{L}) or Target (\texttt{R}).\\
		Target sequences & \texttt{1}-\dots-\texttt{A}-\texttt{X} or \texttt{2}-\dots-\texttt{B}-\texttt{Y}.\\
		& 	Probability of target sequence is $25\%$.\\			
		Training dataset &  Sequence of outer loops starting with \texttt{1} or \texttt{2}.\\ 
		& Maximum number of training samples is $1000000$.\\	
		Inner loops &  Each outer loop contains a random number of inner loops,\\
        & between $1$ and $4$\\
		\bottomrule
	\end{tabular}
\end{table*}

We studied the performance of the \AuGMEnT{} network \citep{AuGMEnT} and our hybrid variant on the sequence prediction task. The mean trend of the RPE-based energy function defined in equation \eqref{AuGMEnT_functional} (Fig. \ref{fig:distr_analysis}A) shows that both models converge in a few hundreds of iterations. As a control, we also simulated a variant in which also memory units were leaky. We noticed that hybrid and standard \AuGMEnT{} networks are more efficient than the purely leaky control. This is not surprising because the key point in the sequence prediction task consists in the maintenance of the initial stimulus, which is simpler with a non-leaky memory than with a leaky one. We notice that the hybrid model has a behaviour similar to \AuGMEnT{}.\\ 

We also analyzed the effect of the temporal length of the sequences on the network performance, by varying the number of distractors (i.e. the intermediate letters) per sequence (Fig. \ref{fig:distr_analysis}B). For each sequence length, the network was retrained ab initio. We required $100$ consecutive correct predictions as the criterion for convergence. We ran $100$ simulations starting with different initializations for each sequence length and averaged the convergence time. Again, \AuGMEnT{} and Hybrid \AuGMEnT{} show good learning performance, maintaining an average of about $250$ trials to convergence for sequences containing up to $10$ distractors, whereas a network with purely leaky units is much slower to converge.\\

The leaky dynamics are not helpful for the sequence prediction task, because the intermediate cues are not relevant for the final model performance. Therefore, it is sufficient to supress the weight values in the $\textbf{V}^{M}$ matrix for distractors, and increase those of the initial \texttt{A}/\texttt{X} letter. This is confirmed by the structure of the weight matrices of the memory in all networks shown after convergence (Fig. \ref{fig:weight_analysis}) in simulations of the sequence prediction task on sequences of five and ten letters. The weight values are highest in absolute value for letters \texttt{A} and \texttt{X}, for both the ON ($+$) and OFF ($-$) units. Finally, we also notice that Hybrid \AuGMEnT{} employs mainly the conservative or non-leaky memory units (\texttt{M1-C} and \texttt{M2-C}) rather than the leaky ones (\texttt{M1-L} and \texttt{M2-L}) to solve the task, showing that the network is able to focus the update dynamics on the connections that are more suitable for the specific task.\\

\subsection{Task 2: 12AX}	
\label{12AX}
The 12AX task is a standard cognitive task used to test Working Memory and diagnose behavioral and cognitive deficits related to memory dysfunctions \citep{HER2015}. Basically, the problem consists in identifying some target sequences among a group of symbols that appear on a screen.\\

The general procedure of the task is schematized in Figure \ref{fig:AugMEnT}A and details involving the construction of the 12AX dataset are collected in Table \ref{tab:12AX_param}. The set of possible stimuli consists of $8$ symbols: two digit cues (\texttt{1} and \texttt{2}), two context cues (\texttt{A} and \texttt{B}), two target cues (\texttt{X} and \texttt{Y}) and finally two distractors (\texttt{C} and \texttt{Z}). Each trial (or outer loop) starts with a digit cue and is followed by a random number of $1$ to $4$ inner loops. Inner loops are composed of patterns of context-target cues, like \texttt{A}-\texttt{X}, \texttt{B}-\texttt{X} or \texttt{B}-\texttt{Y}. The distractors are non task-relevant cues that can invalidate a subsequence creating wrong inner loops like \texttt{A}-\texttt{Z} or \texttt{C}-\texttt{X}. The cues are presented one by one on a screen and the agent has two possible responses for each of them: Target (\texttt{R}) and Non-Target (\texttt{L}). There are only two valid Target cases: in trials that start with digit \texttt{1}, the Target is associated with the target cue \texttt{X} if preceded by context \texttt{A} (\texttt{1}-\dots-\texttt{A}-\texttt{X}); otherwise, in case of initial digit \texttt{2}, the Target occurs if the target cue \texttt{Y} comes after context \texttt{B} (\texttt{2}-\dots-\texttt{B}-\texttt{Y}). The dots are inserted to stress that the target inner loop can occur even a long time after the digit cue, as happens in the following example sequence:  \textbf{\texttt{1}}-\texttt{A}-\texttt{Z}-\texttt{B}-\texttt{Y}-\texttt{C}-\texttt{X}-\textbf{\texttt{A}}-\textbf{\texttt{X}} (whose sequence of correct responses is \texttt{L}-\texttt{L}-\texttt{L}-\texttt{L}-\texttt{L}-\texttt{L}-\texttt{L}-\texttt{L}-\textbf{\texttt{R}}). The variability in the temporal length of each trial is the main issue in solving the 12AX task because of the temporal credit assignment problem. Moreover, since \texttt{1}-\texttt{A}-\texttt{X} and \texttt{2}-\texttt{B}-\texttt{Y} are target sequences, whereas \texttt{2}-\texttt{A}-\texttt{X} and \texttt{1}-\texttt{A}-\texttt{Y} are not, the task can be seen as a generalization of temporal XOR (Supplementary Fig. \ref{fig:spatial_tasks}).\\

The types of the inner loops are determined randomly, with a probability of $50\%$ to have pairs \texttt{A}-\texttt{X} or \texttt{B}-\texttt{Y}. As a result, combined with the probability to have either \texttt{1} or \texttt{2} as starting digit of the trial, the overall probability to have target pair is $25\%$. Since the Target response \texttt{R} has to be associated only with an \texttt{X} or \texttt{Y} stimulus that appears in the correct sequence, the number of Non-Targets \texttt{L} is generally much larger, on average $8.96$ Non-Targets to $1$ Target. We rewarded the correct predictions of a Non-Target with $0.1$ and of Targets with $1$, and punished the wrong predictions with reward of $-1$. In effect, we balanced the positive reward approximately equally between Targets and Non-Targets based on their relative frequencies, which aids convergence.\\

\begin{figure*}[h!]
	\centering
	\begin{minipage}[t]{0.03\textwidth}
        \textbf{A}
    \end{minipage}
    \begin{minipage}[t]{0.29\textwidth}
	    \includegraphics[width = \textwidth,valign=t]{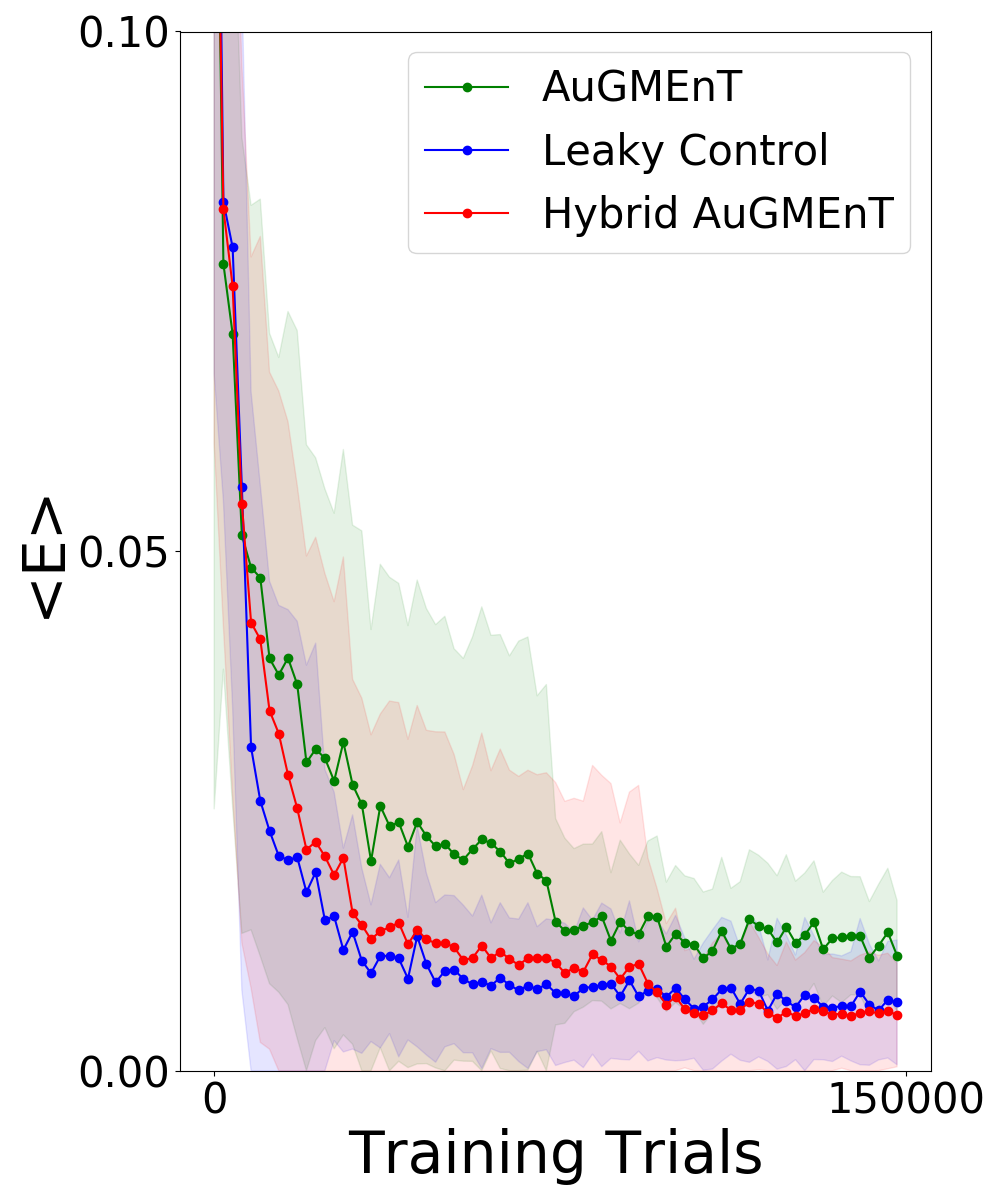}
    \end{minipage}\hfill
	\begin{minipage}[t]{0.03\textwidth}
        \textbf{B}
    \end{minipage}
    \begin{minipage}[t]{0.29\textwidth}
	    \includegraphics[width = \textwidth,valign=t]{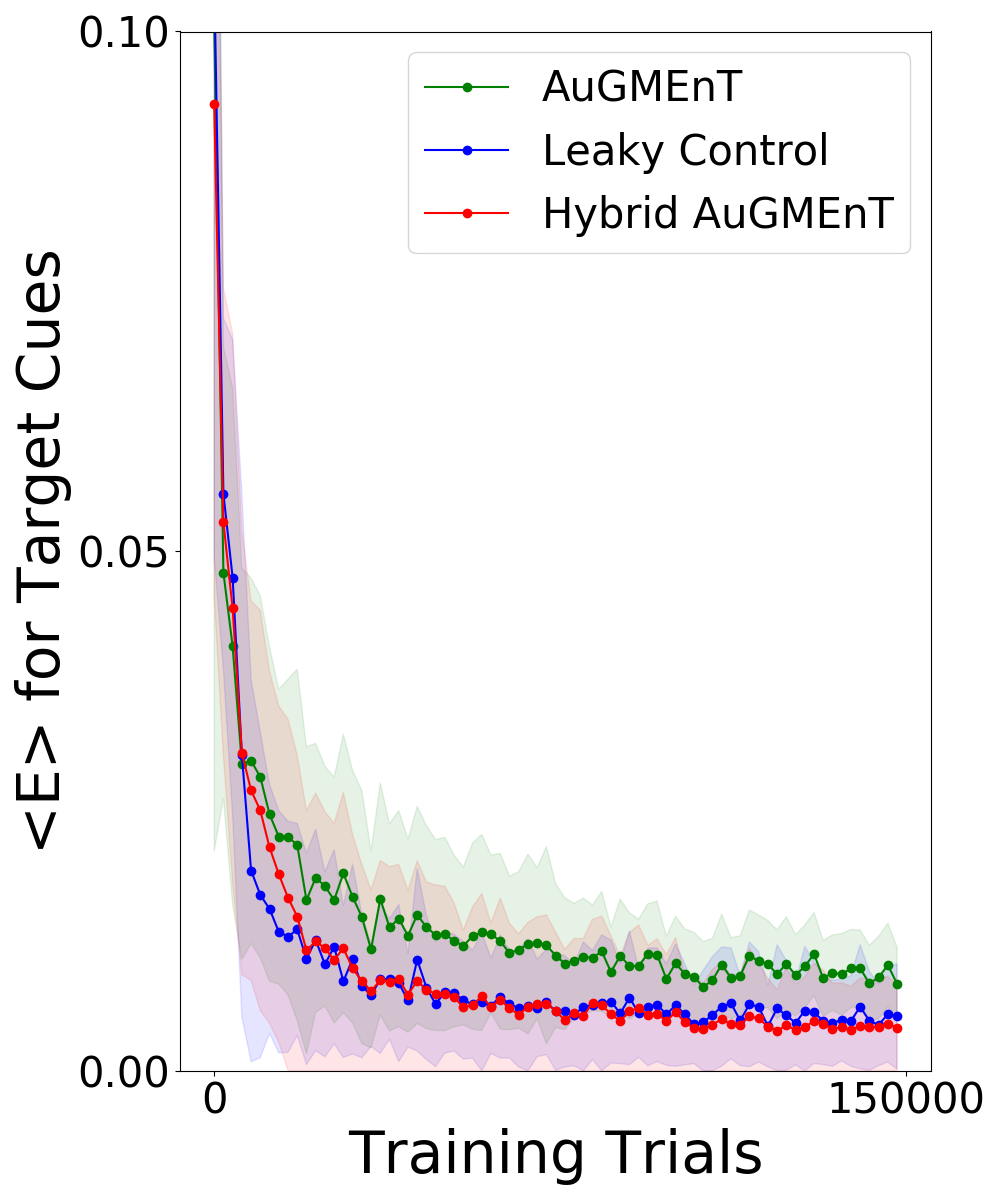}
    \end{minipage}\hfill
	\begin{minipage}[t]{0.03\textwidth}
        \textbf{C}
    \end{minipage}
    \begin{minipage}[t]{0.29\textwidth}
	    \includegraphics[width = \textwidth,valign=t]{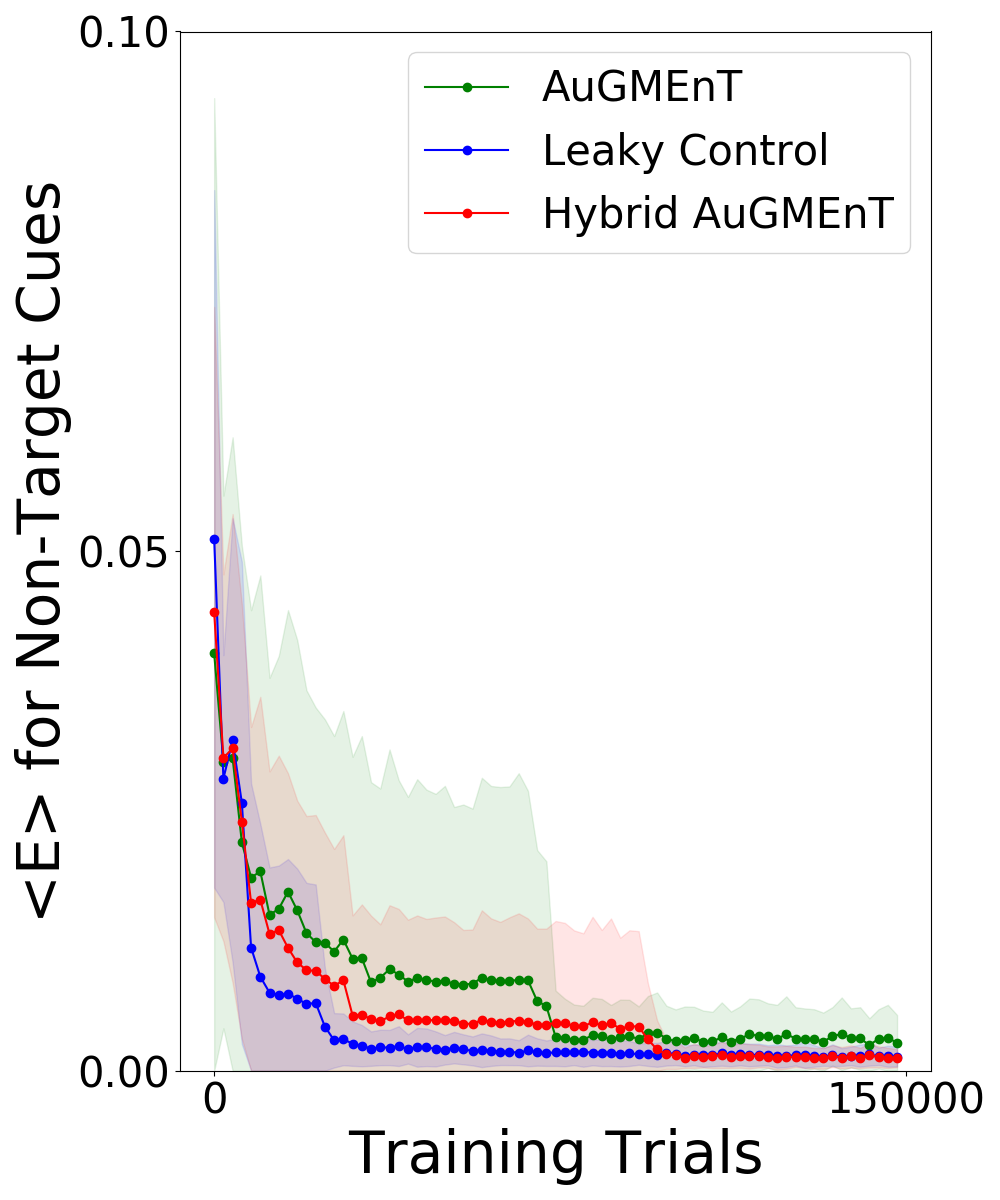}
    \end{minipage}\hfill    
    
	\caption[]{\textbf{Learning convergence of the \AuGMEnT{} variants in the 12AX task.} Minimization of the RPE-based energy function during training on the 12AX task. \textbf{A}. All networks show a good decay of the mean-squared RPE, but they seem to converge to a non-zero regime and, in particular, the base \AuGMEnT{} network (green) is the one that maintains a higher mean-squared RPE level when compared to leaky control (blue) and Hybrid \AuGMEnT{} (red). \textbf{B}. Mean-square RPE associated with only potential target cues \texttt{X} and \texttt{Y}. \textbf{C}. Mean-squared RPE related to only non-target cues.}
	\label{fig:12AX_conv_analysis}
\end{figure*}
\begin{figure*}[h!]
	\centering
	\includegraphics[width=0.95\textwidth]{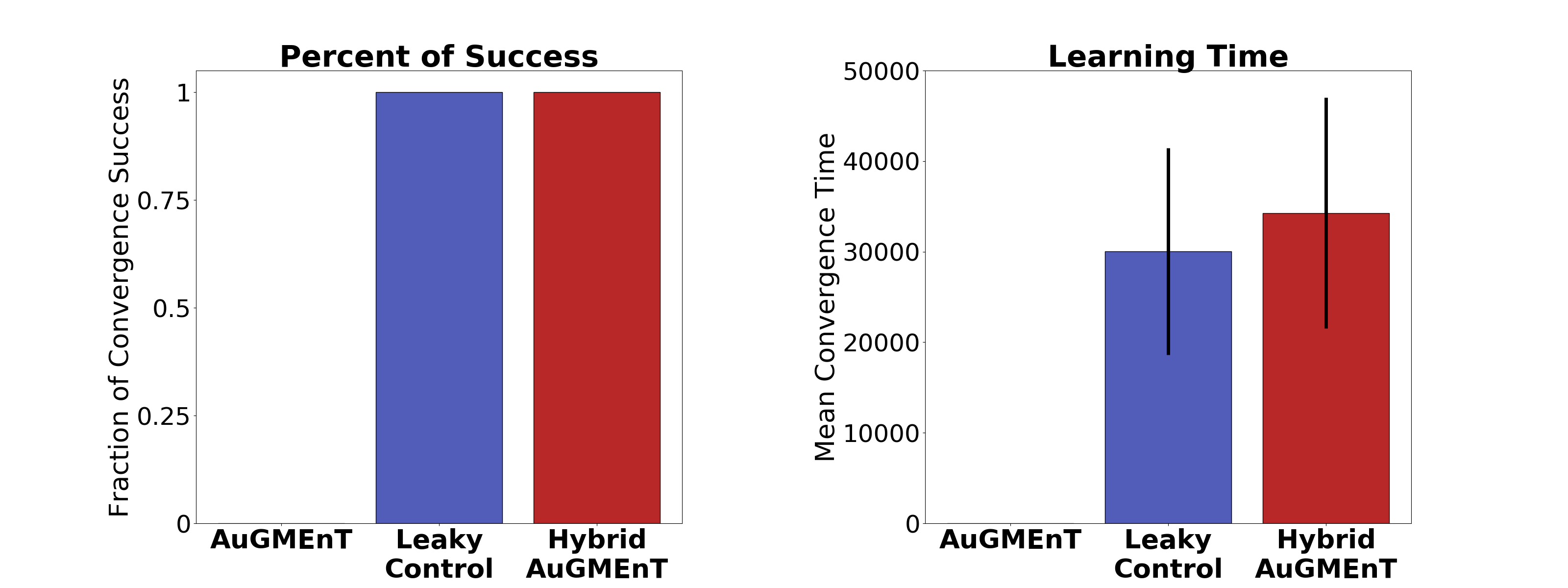}
	\caption[]{\textbf{Comparative statistics of the \AuGMEnT{} variants on performance on the 12AX task.} Barplot description of the learning behavior of the three networks  on the 12AX task according to the convergence criterion given by \cite{HER2015}. After $100$ simulations, we measured the fraction of times that the model satisfies convergence condition (left) and the average number of training trials needed to meet the same convergence criterion (right). Although training dataset consists of $1,000,000$ outer loops, the base \AuGMEnT{} network never manages to satisfy the convergence criterion, while the leaky (blue) and hybrid (red) models have similar convergence performance with a learning time of about $30,000$ trials.}
	\label{fig:12AX_barplots}
\end{figure*}
\begin{figure*}[h!]
	\centering
	\includegraphics[height=0.4\textheight]{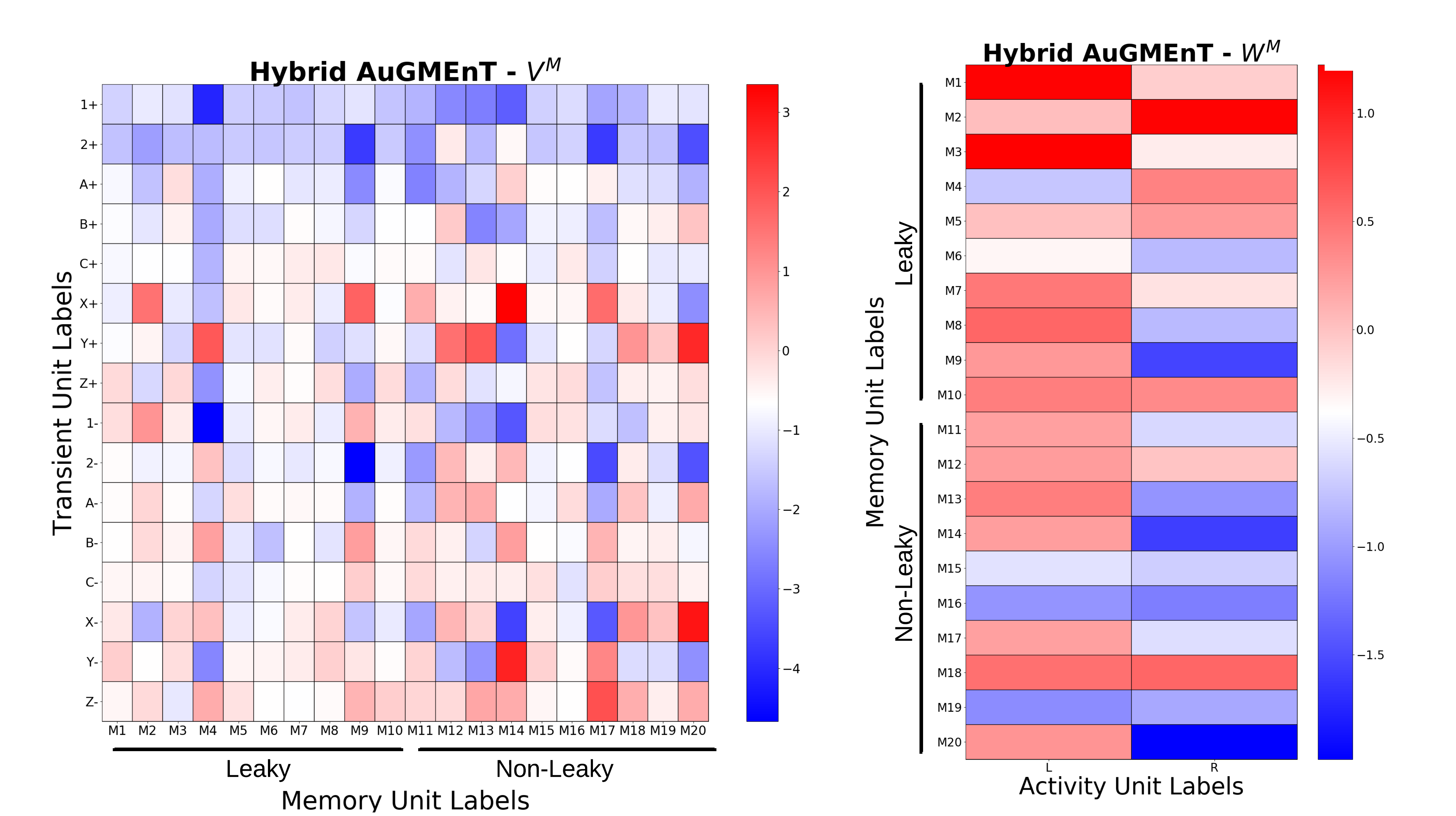}
	\caption[]{\textbf{Memory weights of Hybrid \AuGMEnT{} in the 12AX task.} Plot of the weight matrices in the memory branch of hybrid \AuGMEnT{} network after convergence on the 12AX Task. Left: weights from the transient stimulus into the $20$ memory units (half leaky, half conservative). Right: weights from the memory cells into the output units.}
	\label{fig:12AX_weight_analysis}
\end{figure*}

We simulated Hybrid \AuGMEnT{} network, base \AuGMEnT{} and leaky control on the 12AX task, in order to see whether in this case the introduction of the leaky dynamics improves learning performance. Figure \ref{fig:12AX_conv_analysis}A shows the evolution of the mean squared RPE for the three networks. After a sharp descent, all networks converge to an error level that is non-zero, indicating that learning of the 12AX task is not completely achieved, possibly due to memory interference. However, hybrid \AuGMEnT{} and leaky control saturate at a lower error value than base \AuGMEnT. This difference can be attributed mostly to the errors in responding to the Target cues (Fig. \ref{fig:12AX_conv_analysis}B) than to Non-Target cues (Fig. \ref{fig:12AX_conv_analysis}C). Note that, since 12AX is a Continuous Performance Task (CPT), the error is computed at each iteration -- including the more frequent and trivial Non-Target predictions -- and averaged over $2,000$ consecutive predictions. All networks quickly learn to recognize the Non-Target cues (\texttt{1}, \texttt{2}, \texttt{A}, \texttt{B}, \texttt{C}, \texttt{Z} are always Non-Targets) (Fig. \ref{fig:12AX_conv_analysis}C). However, hybrid \AuGMEnT{} and leaky control learn the more complex identification of Target patterns within a trial when \texttt{X} or \texttt{Y} are presented to the network, better than base \AuGMEnT{} (Fig. \ref{fig:12AX_conv_analysis}B). The gap in the mean squared RPE between hybrid \AuGMEnT{} and leaky control versus base \AuGMEnT{} is wider when only potential target cues are considered in the mean-squared RPE as in Figure \ref{fig:12AX_conv_analysis}B, than when only non-targets are considered as in Figure \ref{fig:12AX_conv_analysis}C.\\

Adopting the convergence criterion from \cite{HER2015} that requires $1,000$ consecutive correct predictions, we show the percentage of successful learning over $100$ simulations and the average learning time in Figure \ref{fig:12AX_barplots} for the three networks. Standard \AuGMEnT{} network was unable to match the convergence condition during any simulation ($0\%$ success), despite presenting 1,000,000 outer loop trials in each simulation. However, hybrid \AuGMEnT{} and leaky control performed 100\% consistently, suggesting that leaky memory units are necessary for the 12AX task. Leaky control learned slightly faster (learning time mean=$30,032.2$ and s.d.=$11,408.9$) than hybrid \AuGMEnT{} (learning time mean=$34,263.6$ and s.d.=$12,737.3$).\\

In order to understand how the hybrid memory works on the 12AX task, we analyzed the weight structure of the connectivity matrices which belong to the memory branch of the hybrid \AuGMEnT{} network (Figure \ref{fig:12AX_weight_analysis}). Unlike in the sequence prediction task, here the hybrid network employs both the leaky and the non-leaky memory units. However, there is an overall separation in the memory activity between the two groups of cells (Fig. \ref{fig:12AX_weight_analysis}, left panel): the leaky units are mainly responsible for the storage of the digit information, having the highest values in absolute value on the weights associated with \texttt{1}($\pm$) and \texttt{2}($\pm$) (e.g. on \texttt{M4} and \texttt{M9}), while the non-leaky cells emphasize more the information coming from the potential Target cues \texttt{X}($\pm$) and \texttt{Y}($\pm$) (e.g. on \texttt{M14}, \texttt{M17} and \texttt{M20}). The storage of the initial digit cue is 'assigned' to leaky units, because the interference from the following letters is reduced thanks to the gradual loss of information while the digit information can survive through sufficient increasing of the digit-related input weights. In this way, the memory interference problem is mitigated, because the crucial digit information is maintained over time without interference in the leaky units and the identification of the inner loops is done by the conservative part of the memory. As a result, all memory units contribute to the definition of the activity Q-values (Fig. \ref{fig:12AX_weight_analysis}, right panel) and, in particular, the memory units that are more active (i.e. the same ones mentioned above) are the ones that strongly discriminate Non-Targets (\texttt{L}) against Targets (\texttt{R}), giving positive contribution to one and negative to the other.\\

The memory units show an opposing behavior on activation versus on deactivation of Target cues: for instance, if \texttt{X+} has strong positive weight intensity, then \texttt{X-} shows a contrary negative weight intensity (see \texttt{M14} or \texttt{M17}). In this way, the network tries to reduce the problems of memory interference between subsequent inner cycles by adding to the memory during deactivation, an opposite amount of information stored during the previous activation, effectively erasing the memory. Further, the difference in absolute value between activation and deactivation is higher in case of the leaky cells, because the deactivation at the next iteration has to remove only a lower amount of information from the memory due to leakage. However, for the digit cues \texttt{1} and \texttt{2}, the weights for activation and deactivation have typically the same sign in order to reinforce the digit signal in memory in two subsequent timesteps (see on \texttt{M4} and \texttt{M9}).\\

In conclusion, the conservative dynamics of the memory in standard \AuGMEnT{} can be a limitation for the learning ability of the model, especially in cases of complex tasks with many data to store or long trials. In fact, even though the complexity of the 12AX task is limited compared to other typical RL tasks, the \AuGMEnT{} network fails to maintain a sufficiently stable performance to satisfy the required convergence criterion. The introduction of the leaky co-efficient in Hybrid \AuGMEnT{} leads to the network solving the 12AX task, overcoming memory interference. However, the loss of information from the leaky memory does not improve learning in other tasks with lower risk of memory interference like the sequence prediction task. Hybrid \AuGMEnT can be adapted to different task structures and to different temporal scales by varying the size and the composition of the memory, for example by considering multiple subpopulations of neurons with distinct memory timescales, say in a power law distribution.\\
 
\section{Discussion}
\label{discussion}

A key goal of the computational neuroscience community is to develop neural networks that are at the same time biologically plausible and able to learn complex tasks similar to humans. The embedding of memory is certainly an important step in this direction, because memory plays a central role in human learning and decision making. Our interest in the \AuGMEnT{} network \citep{AuGMEnT} derives mainly from the biological plausibility of its learning and memory dynamics. In particular, the biological setting of the learning algorithm is based on synaptic tagging, attentional feedback and neuromodulation, providing a possible biological interpretation to backpropagation-like methods.\\

We developed Hybrid \AuGMEnT{}, by introducing leaky dynamics in the memory system, with the aim of improving its learning performance and extending the variety of solvable tasks.
Hybrid \AuGMEnT{} with both leaky and non-leaky units in its memory system, solves the 12AX task on which standard \AuGMEnT{} fails. Both solve the simpler saccade-antisaccade and sequence prediction tasks. Hybrid \AuGMEnT{} inherits the biological plausibility of base \AuGMEnT{} \citep{AuGMEnT}. In addition, consistent learning with decaying memory units requires that the decay of synaptic traces in a memory unit, as per equation \eqref{AuGMEnT_syntrace} \citep{pfister,morrison}, be at the same timescale as decay of the unit's memory state as per equation \eqref{AuGMEnT_cumulative_memory}.\\

Despite the improvement with our hybrid variant, the learning ability of \AuGMEnT{} is still limited compared to other state-of-the-art memory-augmented networks. For instance, the Hierarchical Temporal Memory (\texttt{HTM}) network \citep{seq_pred} presents a greater flexibility in sequence learning than what we have experienced in \AuGMEnT{} on the simple sequence prediction task. Utilizing a complex column-based architecture and an efficient system of inner inhibitions, the \texttt{HTM} network is able to maintain a dual neural activity, both at column level and at unit level, that allows to have sparse representations of the input and give multi-order predictions using an unsupervised Hebbian-like learning rule. 
Nonetheless, it is unclear how the \texttt{HTM} network can be applied to reward-based learning, in particular to tasks like the 12AX, with variable number of inner loops.\\

Although the hybrid memory in the \AuGMEnT{} network remarkably improved its convergence performance on the 12AX task, its learning efficiency is still lower than the reference Hierarchical Error Representation model (\HER{}) \citep{HER2015,HER2016}. In fact, in our simulations, hybrid \AuGMEnT{} showed a mean time to convergence equal to $34,263.6$ outer loops, while the average learning time of \HER{} on the same convergence condition is around $750$ outer loops. The reason for this large gap in the learning performance resides in the gating mechanism of \HER{} network that is specifically developed for hierarchical tasks and is used to decide at each iteration whether to store the new input or maintain the previous content in memory. Unlike \HER{} model, the memory in \AuGMEnT{} does not include any gating mechanism, meaning that the network does not learn when to store and recall information but the memory dynamics are entirely developed via standard weight modulation. On the other hand, the \HER{} model is not as biologically plausible as the \AuGMEnT{} network, because, although its hierarchical structure is inspired on the supposed organization of the prefrontal cortex, its learning scheme is artificial and based on standard backpropagation.\\

In addition, the recent delta-\texttt{RNN} network \citep{deltaRNN} presents interesting similarities with hybrid \AuGMEnT{} in employing two timescales, maintaining memory via interpolation of fast and slow changing inner representations. 
The delta-\texttt{RNN}, whose memory dynamics are a generalization of the gating mechanisms of \LSTM{} and \texttt{GRU}, outperforms these popular recurrent architectures. Thus, it likely has a better learning ability than hybrid \AuGMEnT{}, though it requires a higher number of parameters and the network is not based on biological considerations.\\

The lack of a memory gating system is a great limitation for \AuGMEnT{} variants, when compared with networks equipped with a gated memory, like \HER{} \citep{HER2015,HER2016} or \LSTM{} \citep{LSTM,LSTM2}, especially on complex tasks with high memory demand. Still, even though it cannot be properly defined as a gating system, the forgetting dynamics introduced in hybrid \AuGMEnT{} has a similar effect as the activity of the forget gates in \LSTM{} or \texttt{GRU}. However, unlike forget gates, the decay coefficients are not learnable and are not input-dependent for each memory cell. The Hybrid \AuGMEnT{} network could be further developed by adding a gating control on the leakage: leak gates could be an output of the controller branch of the network and then applied as a gate or decay co-efficient in the memory branch. In this way, the gating value becomes stimulus-dependent and leakage is adjusted to optimize the model performance. On the other hand, such a gating system would make the network more complex, where learning of the gate variables implies an error backpropagation through multiple layers, that may compromise the biological plausibility of the \AuGMEnT{} learning dynamics (though see \citep{Lillicrap,baldi,Lillicrap2}).\\

Alternatively, inspired by the hierarchical architecture of \HER{} \citep{HER2015}, the memory in \AuGMEnT{} could be divided into multiple levels each with their own memory dynamics: each memory level could be associated with distinct synaptic decay and leaky coefficients, learning rates, or gates, in order to cover different temporal scales and encourage level specialization. Compared with hybrid \AuGMEnT{}, this differentiation in memory will not only involve the leaky dynamics, but also the temporal dynamics associated with attentional feedback and synaptic potentiation.\\ 

In the past years, the reinforcement learning community has proposed several deep RL networks, like deep Q-networks \citep{mnih2015human} or the AlphaGo model \citep{alphago}, that combine the learning advantages of deep neural networks with reinforcement learning \citep{li2017deep}. Thus, it may be interesting to consider a deep version of the \AuGMEnT{} network with additional hidden layers of neurons. While conventional error backpropagation in \AuGMEnT{} may not yield local synaptic plasticity, locality might be retained with alternative backpropagation methods \citep{Lillicrap, baldi,Lillicrap2}.\\

\section{Acknowledgements}
We thank Vineet Jain for helpful discussions. Financial support was provided by the European Research Council (Multirules, grant agreement no. 268689), the Swiss National Science Foundation (Sinergia, grant agreement no. CRSII2\_147636), and the European Commission Horizon 2020 Framework Program (H2020) (Human Brain Project, grant agreement no. 720270).\\

\bibliography{AuGMEnT_paper}

\begin{thebibliography}{}

\bibitem[Abbott et~al., 2016]{XOR}
Abbott, L., De~Pasquale, B., and Memmesheimer, R.-M. (2016).
\newblock Building functional networks of spiking model neurons.
\newblock {\em Nature neuroscience}, 19(3):350--355.

\bibitem[Alexander and Brown, 2015]{HER2015}
Alexander, W.~H. and Brown, J.~W. (2015).
\newblock Hierarchical error representation: A computational model of anterior
  cingulate and dorsolateral prefrontal cortex.
\newblock {\em Neural Computation}, 27(11):2354--2410.

\bibitem[Alexander and Brown, 2016]{HER2016}
Alexander, W.~H. and Brown, J.~W. (2016).
\newblock Frontal cortex function derives from hierarchical predictive coding.
\newblock {\em bioRxiv}, page 076505.

\bibitem[Baldi et~al., 2016]{baldi}
Baldi, P., Sadowski, P., and Lu, Z. (2016).
\newblock Learning in the machine: Random backpropagation and the learning
  channel.
\newblock {\em arXiv preprint arXiv:1612.02734}.

\bibitem[Barak and Tsodyks, 2014]{barak}
Barak, O. and Tsodyks, M. (2014).
\newblock Working models of working memory.
\newblock {\em Current opinion in neurobiology}, 25:20--24.

\bibitem[Brzosko et~al., 2015]{brzosko1}
Brzosko, Z., Schultz, W., and Paulsen, O. (2015).
\newblock Retroactive modulation of spike timing-dependent plasticity by
  dopamine.
\newblock {\em eLife}, page 4:e09685.

\bibitem[Brzosko et~al., 2017]{brzosko2}
Brzosko, Z., Zannone, S., Schultz, W., Clopath, C., and Paulsen, O. (2017).
\newblock Sequential neuromodulation of hebbian plasticity offers mechanism for
  effective reward-based navigation.
\newblock {\em eLife}, page 6:e27756.

\bibitem[Chen, 2016]{alphago}
Chen, J.~X. (2016).
\newblock The evolution of computing: Alphago.
\newblock {\em Computing in Science \& Engineering}, 18(4):4--7.

\bibitem[Cho, 2014]{GRU}
Cho, K. e.~a. (2014).
\newblock Learning phrase representations using rnn encoder-decoder for
  statistical machine translation.
\newblock {\em arXiv preprint arXiv:1406.1078}.

\bibitem[Compte et~al., 2000]{compte}
Compte, A., Brunel, N., Goldman-Rakic, P.~S., and Wang, X.-J. (2000).
\newblock Synaptic mechanisms and network dynamics underlying spatial working
  memory in a cortical network model.
\newblock {\em Cerebral Cortex}, 10:910--923.

\bibitem[Cui et~al., 2015]{seq_pred}
Cui, Y., Surpur, C., Ahmad, S., and Hawkins, J. (2015).
\newblock Continuous online sequence learning with an unsupervised neural
  network model.
\newblock {\em CoRR}, abs/1512.05463.

\bibitem[Frank et~al., 2001]{frank}
Frank, M.~J., Loughry, B., and O’Reilly, R.~C. (2001).
\newblock Interactions between frontal cortex and basal ganglia in working
  memory: a computational model.
\newblock {\em Cognitive, Affective, \& Behavioral Neuroscience},
  1(2):137--160.

\bibitem[Fr\'emaux and Gerstner, 2016]{fremaux}
Fr\'emaux, N. and Gerstner, W. (2016).
\newblock Neuromodulated spike-timing-dependent plasticity, and theory of
  three-factor learning rules.
\newblock {\em Frontiers in Neural Circuits}, page 9:85.

\bibitem[Gers and Schmidhuber, 2001]{LSTM2}
Gers, F.~A. and Schmidhuber, J. (2001).
\newblock Long short-term memory learns context free and context sensitive
  languages.
\newblock In {\em Artificial Neural Nets and Genetic Algorithms}, pages
  134--137. Springer.

\bibitem[Gottlieb and Goldberg, 1999]{gottlieb1999}
Gottlieb, J. and Goldberg, M.~E. (1999).
\newblock Activity of neurons in the lateral intraparietal area of the monkey
  during an antisaccade task.
\newblock {\em Nature neuroscience}, 2(10):906--912.

\bibitem[Graves et~al., 2014]{NTM}
Graves, A., Wayne, G., and Danihelka, I. (2014).
\newblock Neural turing machines.
\newblock {\em arXiv preprint arXiv:1410.5401}.

\bibitem[Graves et~al., 2016]{DNC}
Graves, A., Wayne, G., Reynolds, M., Harley, T., Danihelka, I.,
  Grabska-Barwi{\'n}ska, A., Colmenarejo, S.~G., Grefenstette, E., Ramalho, T.,
  Agapiou, J., et~al. (2016).
\newblock Hybrid computing using a neural network with dynamic external memory.
\newblock {\em Nature}, 538(7626):471--476.

\bibitem[Guerguiev et~al., 2017]{Lillicrap2}
Guerguiev, J., Lillicrap, T.~P., and Richards, B.~A. (2017).
\newblock Towards deep learning with segregated dendrites.
\newblock {\em eLife}, page 6:e22901.

\bibitem[He et~al., 2015]{he}
He, K., Huertas, M., Hong, S.~Z., Tie, X., Hell, J.~W., Shouval, H., and
  Kirkwood, A. (2015).
\newblock Distinct eligibility traces for ltp and ltd in cortical synapses.
\newblock {\em Neuron}, 88(3):528–538.

\bibitem[Hochreiter and Schmidhuber, 1997]{LSTM}
Hochreiter, S. and Schmidhuber, J. (1997).
\newblock Long short-term memory.
\newblock {\em Neural computation}, 9(8):1735--1780.

\bibitem[Legenstein et~al., 2008]{legenstein}
Legenstein, R., Pecevski, D., and Wolfgang, M. (2008).
\newblock A learning theory for reward-modulated spike-timing-dependent
  plasticity with application to biofeedback.
\newblock {\em PLOS Comput Biol.}, 4(10):e1000180.

\bibitem[Li, 2017]{li2017deep}
Li, Y. (2017).
\newblock Deep reinforcement learning: An overview.
\newblock {\em arXiv preprint arXiv:1701.07274}.

\bibitem[Lillicrap et~al., 2016]{Lillicrap}
Lillicrap, T.~P., Cownden, D., Tweed, D.~B., and Akerman, C.~J. (2016).
\newblock Random synaptic feedback weights support error backpropagation for
  deep learning.
\newblock {\em Nature communications}, 7.

\bibitem[Mao et~al., 2011]{mao2011}
Mao, T., Kusefoglu, D., Hooks, B.~M., Huber, D., Petreanu, L., and Svoboda, K.
  (2011).
\newblock Long-range neuronal circuits underlying the interaction between
  sensory and motor cortex.
\newblock {\em Neuron}, 72(1):111--123.

\bibitem[Mink, 1996]{mink}
Mink, J.~W. (1996).
\newblock The basal ganglia: focused selection and inhibition of competing
  motor programs.
\newblock {\em Progress in neurobiology}, 50(4):381--425.

\bibitem[Minsky and Papert, 1969]{xor1}
Minsky, M. and Papert, S. (1969).
\newblock Perceptrons.
\newblock {\em MIT press}.

\bibitem[Mnih et~al., 2015]{mnih2015human}
Mnih, V., Kavukcuoglu, K., Silver, D., Rusu, A.~A., Veness, J., Bellemare,
  M.~G., Graves, A., Riedmiller, M., Fidjeland, A.~K., Ostrovski, G., et~al.
  (2015).
\newblock Human-level control through deep reinforcement learning.
\newblock {\em Nature}, 518(7540):529--533.

\bibitem[Moore and Armstrong, 2003]{gateattention2}
Moore, T. and Armstrong, K.~M. (2003).
\newblock Selective gating of visual signals by microstimulation of frontal
  cortex.
\newblock {\em Nature}, 421(6921):370--373.

\bibitem[Morrison et~al., 2008]{morrison}
Morrison, A., Diesmann, M., and Gerstner, W. (2008).
\newblock Phenomenological models of synaptic plasticity based on spike timing.
\newblock {\em Biological Cybernetics}, 98(6):459--478.

\bibitem[Okano et~al., 2000]{memoryrole}
Okano, H., Hirano, T., and Balaban, E. (2000).
\newblock Learning and memory.
\newblock {\em Proceedings of the National Academy of Sciences},
  97(23):12403--12404.

\bibitem[O'Reilly and Frank, 2006]{PBWM}
O'Reilly, R.~C. and Frank, M.~J. (2006).
\newblock Making working memory work: a computational model of learning in the
  prefrontal cortex and basal ganglia.
\newblock {\em Neural computation}, 18(2):283--328.

\bibitem[Ororbia et~al., 2017]{deltaRNN}
Ororbia, A.~G., Mikolov, T., and Reitter, D. (2017).
\newblock Learning simpler language models with the differential state
  framework.
\newblock {\em Neural Computation}, 29(12):3327--3352.

\bibitem[Pfister and Gerstner, 2006]{pfister}
Pfister, J.-P. and Gerstner, W. (2006).
\newblock Triplets of spikes in a model of spike timing-dependent plasticity.
\newblock {\em Journal of Neuroscience}, 26(38):9673--9682.

\bibitem[Roelfsema and van Ooyen, 2005]{AGREL}
Roelfsema, P.~R. and van Ooyen, A. (2005).
\newblock Attention-gated reinforcement learning of internal representations
  for classification.
\newblock {\em Neural computation}, 17(10):2176--2214.

\bibitem[Roelfsema et~al., 2010]{gateattention1}
Roelfsema, P.~R., van Ooyen, A., and Watanabe, T. (2010).
\newblock Perceptual learning rules based on reinforcers and attention.
\newblock {\em Trends in cognitive sciences}, 14(2):64--71.

\bibitem[Rombouts et~al., 2015]{AuGMEnT}
Rombouts, J.~O., Bohte, S.~M., and Roelfsema, P.~R. (2015).
\newblock How attention can create synaptic tags for the learning of working
  memories in sequential tasks.
\newblock {\em PLOS Computational Biology}, 11(3):1--34.

\bibitem[Rumelhart et~al., 1985]{xor2}
Rumelhart, D.~E., Hinton, G.~E., and Williams, R.~J. (1985).
\newblock Learning internal representations by error propagation.
\newblock Technical report, California Univ San Diego La Jolla Inst for
  Cognitive Science.

\bibitem[Samsonovich and McNaughton, 1997]{samsonovich}
Samsonovich, A. and McNaughton, B.~L. (1997).
\newblock Path integration and cognitive mapping in a continuous attractor
  neural network model.
\newblock {\em Journal of Neuroscience}, 17(15):5900--5920.

\bibitem[Santoro et~al., 2016]{oneshotNTM}
Santoro, A., Bartunov, S., Botvinick, M., Wierstra, D., and Lillicrap, T.
  (2016).
\newblock One-shot learning with memory-augmented neural networks.
\newblock {\em arXiv preprint arXiv:1605.06065}.

\bibitem[Schultz et~al., 1993]{brainsigma1}
Schultz, W., Apicella, P., and Ljungberg, T. (1993).
\newblock Responses of monkey dopamine neurons to reward and conditioned
  stimuli during successive steps of learning a delayed response task.
\newblock {\em Journal of neuroscience}, 13(3):900--913.

\bibitem[Schultz et~al., 1997]{brainsigma2}
Schultz, W., Dayan, P., and Montague, P.~R. (1997).
\newblock A neural substrate of prediction and reward.
\newblock {\em Science}, 275(5306):1593--1599.

\bibitem[Sutton, 1984]{tempproblem}
Sutton, R.~S. (1984).
\newblock {\em Temporal Credit Assignment in Reinforcement Learning}.
\newblock PhD thesis.
\newblock AAI8410337.

\bibitem[Sutton and Barto, 1998]{sutton1998}
Sutton, R.~S. and Barto, A.~G. (1998).
\newblock {\em Reinforcement learning: An introduction}, volume~1.
\newblock MIT press Cambridge.

\bibitem[Tetzlaff et~al., 2012]{intro1}
Tetzlaff, C., Kolodziejski, C., Markelic, I., and W{\"o}rg{\"o}tter, F. (2012).
\newblock Time scales of memory, learning, and plasticity.
\newblock {\em Biological Cybernetics}, 106(11):715--726.

\bibitem[Vasilaki et~al., 2009]{vasilaki}
Vasilaki, E., Fr\'emaux, N., Urbanczik, R., Senn, W., and Gerstner, W. (2009).
\newblock Spike-based reinforcement learning in continuous state and action
  space: When policy gradient methods fail.
\newblock {\em PLOS Computational Biology}, 5(12):e1000586.

\bibitem[Waelti et~al., 2001]{brainsigma3}
Waelti, P., Dickinson, A., and Schultz, W. (2001).
\newblock Dopamine responses comply with basic assumptions of formal learning
  theory.
\newblock {\em Nature}, 412(6842):43--48.

\bibitem[Wiering and Schmidhuber, 1998]{Wiering1998}
Wiering, M. and Schmidhuber, J. (1998).
\newblock Fast online q($\lambda$).
\newblock {\em Machine Learning}, 33(1):105--115.

\bibitem[Xie and Seung, 2004]{xie}
Xie, X. and Seung, H.~S. (2004).
\newblock Learning in neural networks by reinforcement of irregular spiking.
\newblock {\em Phys Rev E.}, 69(4):041909.

\bibitem[Yagishita et~al., 2014]{yagishita}
Yagishita, S., Hayashi-Takagi, A., Ellis-Davies, G.~C., Urakubo, H., Ishii, S.,
  and Kasai, H. (2014).
\newblock A critical time window for dopamine actions on the structural
  plasticity of dendritic spines.
\newblock {\em Science}, 345(6204):1616–1620.

\end{thebibliography}

\clearpage
\setcounter{figure}{0} 
\setcounter{table}{0} 
\twocolumn[{
\begin{center}
	\textbf{\LARGE Supplementary Material}\\[3em]
	\textbf{\Large Model Validation of \AuGMEnT{} Network:\\[0.6em]The Saccade-Antisaccade Task}\\[2em]
\end{center}
}]
\label{SAS}

The Saccade-AntiSaccade Task (S-AS), presented in the \texttt{AuGMEnT} paper \citep{AuGMEnT}, is inspired by cognitive experiments performed on monkeys to study the memory representations of visual stimuli in the Lateral Intra-Parietal cortex (LIP). The structure of each trial covers different phases in which different cues are presented on a screen and at the end of each episode the agent has to respond accordingly in order to gain a reward. Actually, following a shaping strategy, monkeys received also an intermediate smaller reward when they learnt to fixate on the task-relevant marks at the center of the screen. The details on the procedure of the trials and the experimental results are discussed in \cite{gottlieb1999}.\\
\begin{table*}[h!]
	\caption{Table of the S-AS task}
	\label{tab:SAS_param}
	\centering
	\begin{tabular}{l l}
		\toprule
		\textbf{Task feature} & \textbf{Details}  \\
		\midrule
		Task Structure & $5$ phases: \textit{start}, \textit{fix}, \textit{cue}, \textit{delay}, \textit{go} \\			
		Inputs & Fixation mark: Pro-saccade (\texttt{P}) or Anti-saccade (\texttt{A})\\
		& Location mark: Left (\texttt{L}) or Right (\texttt{R})\\
		Outputs & Eye movement: Left (\texttt{L}), Front (\texttt{F}) or Right (\texttt{R})\\	
		Trial Types & 1. \texttt{P}+\texttt{L}=\texttt{L}  $\quad$ 2. \texttt{P}+\texttt{R}=\texttt{R} \\  		 &	3. \texttt{A}+\texttt{L}=\texttt{R} $\quad$ 4. \texttt{A}+\texttt{R}=\texttt{L} \\		
		Training dataset & Maximum number of trials is $25,000$.\\ 
		& Each trial type has equal probability.\\
		Rewards & Correct saccade at \textit{go} ($1.5$ units) \\
		&  Fixation of the screen in \textit{fix} ($0.2$ units) \\					               	
		\bottomrule
	\end{tabular}
\end{table*}
The agent has to look either to 'Left' (\texttt{L}) or to 'Right' (\texttt{R}) in agreement with a sequence of marks that appear on a screen at each episode. The response, corresponding to the direction of the eye movement (called saccade), depends on the combination of the location and fixation marks. The location cue is a circle displayed either at the left side (\texttt{L}) or at the right side (\texttt{R}) of the screen, while the fixation mark is a square presented at the center that indicates whether the final move has to be concordant with the location cue (Prosaccade - \texttt{P}) or in the opposite direction (Antisaccade - \texttt{A}). As a consequence, there are four types of trials corresponding to the four cue combinations.\\ 

\begin{figure}[h!]
	\centering
	\includegraphics[width = 0.45\textwidth]{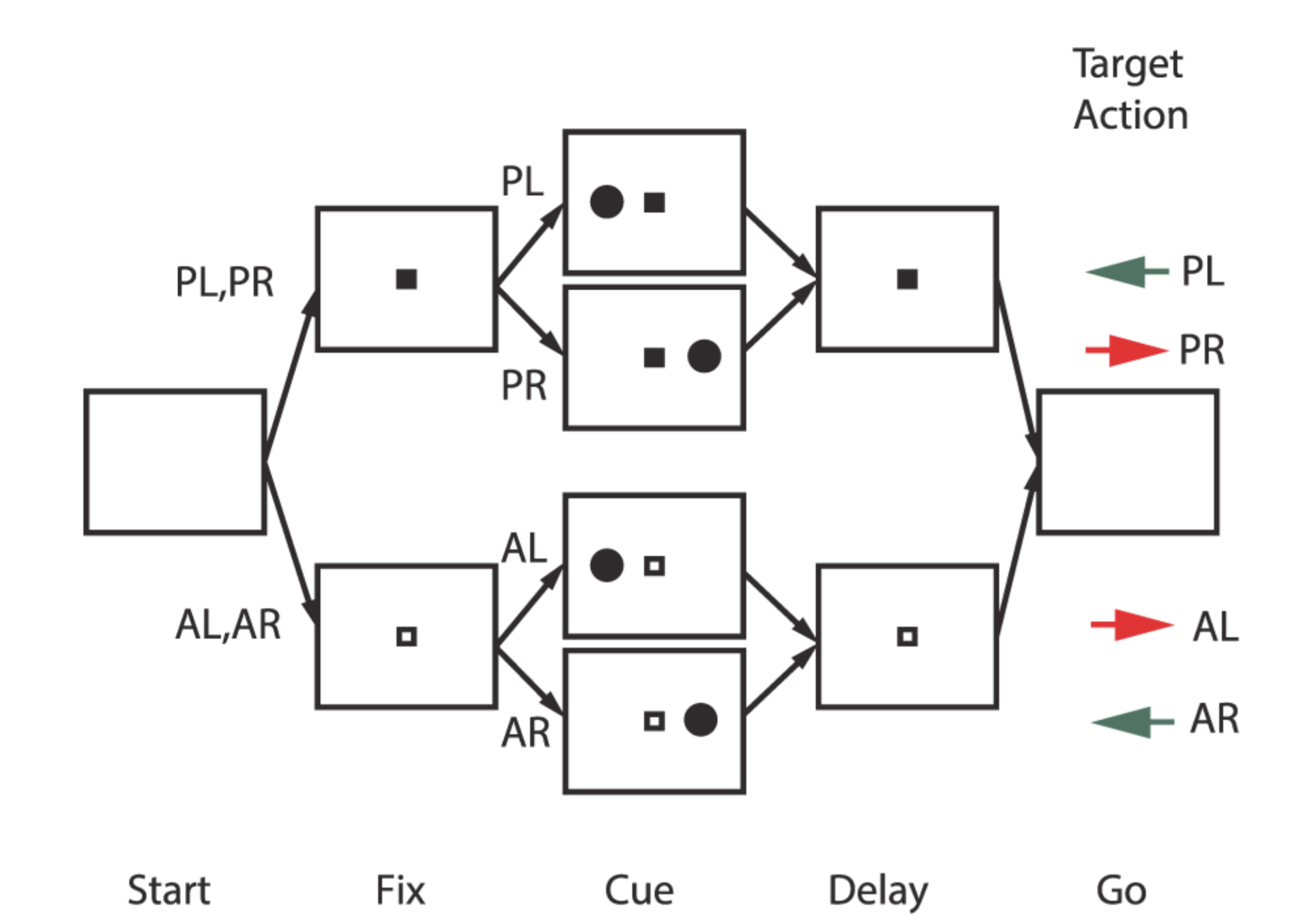}
	\caption[]{\textbf{Structure of the Saccade-AntiSaccade task.} Structure of the trials in all the possible modalities: \texttt{P}-\texttt{L} and \texttt{A}-\texttt{R} have final response \texttt{L} (green arrow), while trials \texttt{P}-\texttt{R} and \texttt{A}-\texttt{L} lead to take action \texttt{R} (red arrow). Figure taken from publication \citep{AuGMEnT}.}
	\label{fig:SAS}
\end{figure}

As can be seen in Figure \ref{fig:SAS}, each trial is structured in five phases: a) \textit{start}, where the screen is initially empty, b) \textit{fix}, when the fixation mark appears c) \textit{cue}, where the location cue is added on the screen, d) \textit{delay}, in which the location circle disappears for two timesteps, e) \textit{go}, when the fixation mark vanishes as well and the agent has to give the final response to get the reward. Since the action is given at the end of the trial when the screen is completely empty, the task can be solved only if the network stores and maintains both the stimuli in memory in spite of the delay phase. In addition, the shaping strategy mentioned above is applied in the \textit{fix} phase of the experiment, by giving an intermediate reward if the agent gazes at the fixation mark for two consecutive timesteps, to ensure that he observes the screen during the whole trial and that the \textit{go} response is not random but consequential to the cues. So, the reward for the final response is equal to $1.5$ units in case of correct response, $0$ otherwise, but the intermediate reward for the shaping strategy is smaller, equal to $0.2$ units. The most important details about the trial structure are summarized in Table \ref{tab:SAS_param}.\\

\begin{figure*}[h!]
	\centering
	\includegraphics[height = 0.32\textheight]{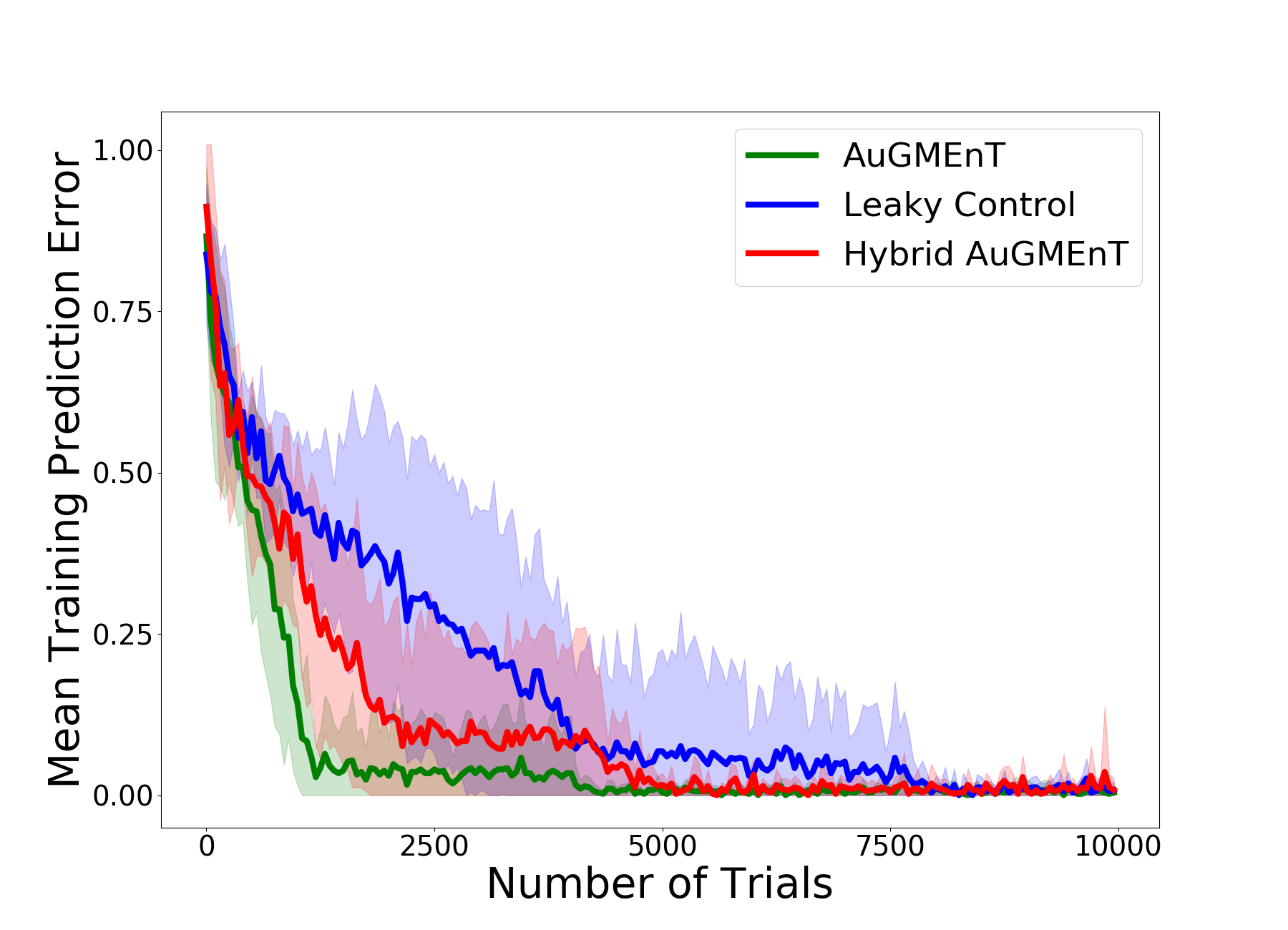}
		\caption[]{\textbf{Validation of learning perfomance on the saccade task.} Decay of the mean predicition error during training of \AuGMEnT{} networks on the S-AS task, computed as the mean number of errors in $50$ trials and averaged over $100$ simulations.}
	\label{fig:SAS_error}
\end{figure*}

The mean trend of the prediction error during training shows that learning of the S-AS task is achieved by the of \AuGMEnT{} network also in our simulation (Figure \ref{fig:SAS_error}). In particular, in order to compare it with the reference performance in \cite{AuGMEnT} we applied the same convergence condition, for which training on the S-AS task is considered to be successful if the accuracy for each trial type is higher than $90\%$ in the last $50$ trials. In the original paper, convergence of \AuGMEnT{} is achieved $99.45\%$ of the times, with a training of around $4,100$ trials. In our simulations, the network reaches convergence every time, with a mean time of $2,063$ trials (s.d.$=837.7$). The slightly better performance in our simulations could be due to minor differences in the interpretation of the task structure of S-AS or of the convergence criterion, but in any case the error plot in Figure \ref{fig:SAS_error} confirms a sharp decrease in the variability of the prediction error after $4,000$ iterations, compliant with the convergence results in the reference paper.\\
In addition, we also show the performance of hybrid \AuGMEnT{} and leaky control, proving that they also solve the S-AS task. However, the time to convergence is higher (especially for leaky control) because non-leaky memory is better suited to this simple task.\\ 

\begin{figure*}[h!]
	\centering
	\begin{minipage}[t]{0.03\textwidth}
        \textbf{A}
    \end{minipage}
    \begin{minipage}[t]{0.22\textwidth}
    	\begin{center}
    	Temporal XOR Task
    	\end{center}
        \bigskip
	    \includegraphics[width = \textwidth,valign=t]{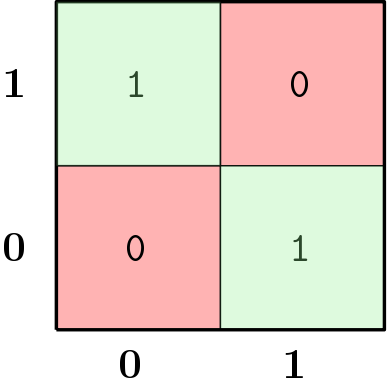}
    \end{minipage}\hfill
   	\begin{minipage}[t]{0.03\textwidth}
        \textbf{B}
    \end{minipage}
    \begin{minipage}[t]{0.22\textwidth}
    	\begin{center}
    	S-AS Task
    	\end{center}
        \bigskip
	    \includegraphics[width = \textwidth,valign=t]{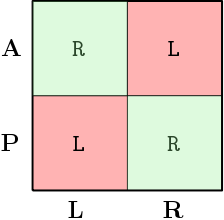}
    \end{minipage}\hfill 
    \begin{minipage}[t]{0.03\textwidth}
        \textbf{C}
    \end{minipage}
    \begin{minipage}[t]{0.4\textwidth}	
    	\begin{center}
    	12AX Task
    	\end{center}
	    \includegraphics[width = \textwidth,valign=t]{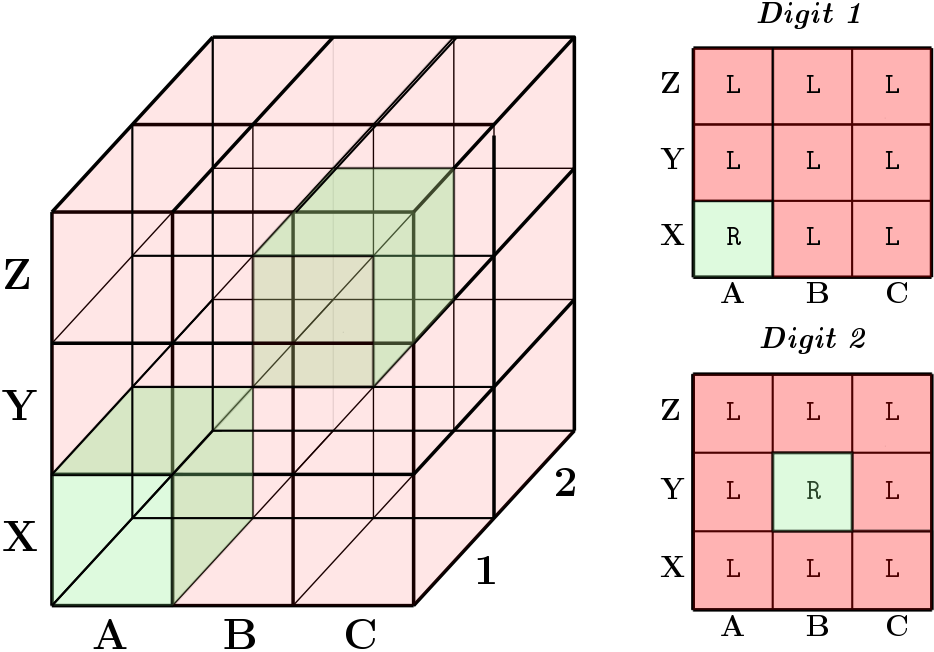}
	\end{minipage}
	\caption[]{\textbf{Spatial representation of temporal XOR, S-AS and 12AX tasks.}  Schematic representation of the task structures indicating for each cue combination the correct response. Temporal XOR (A) and S-AS (B) tasks have analogous input-output maps and they both have an output space that is not linearly separable.  The map of 12AX task (C)  is three-dimensional because its structure is based on three hierarchical levels of inputs instead of two. However, for better visualization we add at right the sections of the structure space with respect to the digit inputs that start the outer loops. The output space is still not linearly separable in 3D.}
	\label{fig:spatial_tasks}
\end{figure*}

\end{document}